\documentclass[11pt,a4paper]{article}
\pdfoutput=1
\usepackage{jheppub}
\usepackage{amsmath,amssymb}
\usepackage{epsfig}
\usepackage{graphics,graphicx}
\usepackage[active]{srcltx}
\usepackage{bbm}

\newcommand{\bea}{\begin{eqnarray}}
\newcommand{\eea}{\end{eqnarray}}
\newcommand{\bean}{\begin{eqnarray*}}
\newcommand{\eean}{\end{eqnarray*}}
\newcommand{\nn}{\nonumber}

\def\W #1{\widetilde{#1}}

\def\braket#1{\left\langle #1 \right\rangle}

\def\gb #1{ \left\langle #1 \right]}

\def\Tr{\mathop{\rm Tr}}
\newcommand{\tr}{\mathop{\rm Tr}}

\def\eref#1{(\ref{#1})}

\def\a{{\alpha}}

\def\eps{\epsilon}

\def\vev{\braket}

\def\bvev#1{\left[ #1 \right]}
\def\Spaa{\vev}
\def\Spbb{\bvev}

\newcommand{\cA}{{\cal A}}

\newcommand{\cV}{{\cal V}}

\newcommand{\IP}{\mathbb{P}}

\newcommand{\IC}{\mathbb{C}}

\newcommand{\IZ}{\mathbb{Z}}

\newtheorem{theorem}{\bf THEOREM}

\newtheorem{corollary}{\bf COROLLARY}

\def\Label#1{\label{#1}%
  \smash{\hbox to0pt{\raise1ex\hbox{\tiny[#1]}\hss}}}

\def\ideal#1{\big\langle #1 \big\rangle}
\def\gb{\mbox{GB}}
\def\B{\mbox{B}}

\def\spaa #1{\langle #1\rangle}
\def\spbb #1{[#1]}
\def\spab #1{\langle #1]}

\def\I{\tiny \mbox{I}}

\title{An Algebraic Approach to the Scattering Equations}
\author[a]{Rijun Huang\footnote{The
unusual ordering of authors is just to let authors get proper
recognition of contributions under outdated practice in China. },}
\author[a]{Junjie Rao,}
\author[a,b]{Bo Feng,}
\author[c,d,e]{and Yang-Hui He}
\affiliation[a]{Zhejiang Institute of Modern Physics,
  Zhejiang University, Hangzhou, 310027, P.R. China}
\affiliation[b]{Center of Mathematical Science,
  Zhejiang University, Hangzhou, 310027, P.R. China}
\affiliation[c]{School of Physics, NanKai University, Tianjin, 300071, P.R. China}
\affiliation[d]{Department of Mathematics, City University, London, EC1V 0HB, UK}
\affiliation[e]{Merton College, University of Oxford, OX14JD, UK}

\emailAdd{huang@nbi.dk}
\emailAdd{raojunjie@zju.edu.cn}
\emailAdd{fengbo@zju.edu.cn}
\emailAdd{hey@maths.ox.ac.uk}

\date{\today}
\abstract{We employ the so-called companion matrix method from computational algebraic geometry, tailored for zero-dimensional ideals, to study the scattering equations.
  The method renders the CHY-integrand of scattering amplitudes computable using simple linear algebra and is amenable to an algorithmic approach.
  Certain identities in the amplitudes as well as rationality of the final integrand become immediate in this formalism.
}


\begin{document}
\maketitle \flushbottom

\section{Introduction}

In last couple of years, amazing progress has been made by Cachazo,
He and Yuan [CHY] in a series of papers
\cite{Cachazo:2013gna,Cachazo:2013hca, Cachazo:2013iea,
Cachazo:2014nsa,Cachazo:2014xea}, where tree-level amplitudes of a
host of quantum field theories can be calculated using solutions of
a set of algebraic equations. These are called the {\bf scattering
equations} and appear in the literature in a variety of contexts
\cite{Fairlie:1972abc,Roberts:1972abc,Fairlie:2008dg,Gross:1987ar,Witten:2004cp,Caputa:2011zk,Caputa:2012pi,Makeenko:2011dm,Cachazo:2012uq}.

The mysterious relationship between the CHY approach and the
standard QFT paradigm has been explained from different points of
view. In \cite{Dolan:2013isa}, using the BCFW on-shell recursion
relation \cite{Britto:2004ap,Britto:2005fq} the validity of the CHY
construction for $\phi^3$ theory and Yang-Mills theories has been
proven. A broader understanding is achieved using ambitwistor string
theory
\cite{Mason:2013sva,Berkovits:2013xba,Gomez:2013wza,Adamo:2013tsa,Geyer:2014fka,
Geyer:2014lca,Casali:2014hfa,Adamo:2015hoa,Casali:2015vta,Ohmori:2015sha,Geyer:2015bja},
where using different world-sheet fields, different integrands in
the CHY approach for different theories -- which we will call {\bf
CHY-integrands}, a function of the coordinates $z_i$ in a Riemann
surface -- have been derived alongside with the natural appearance
of  scattering equations. A nice point of ambitwistor approach is
that it provides the natural framework for loop scattering equations
as studied in  \cite{Adamo:2013tsa,Casali:2014hfa}, which lead to  a
breakthrough in \cite{Geyer:2015bja}. A third understanding is given
in \cite{Bjerrum-Bohr:2014qwa}, where inspired by the field theory
limit of string theory, a dual model has been introduced, based on
which a direct connection between the CHY approach and the standard
Feynman diagram method has been established not only at the
tree-level in \cite{Baadsgaard:2015voa, Baadsgaard:2015ifa}, but
also at the one-loop level (at least for $\phi^3$ theory) in
\cite{Baadsgaard:2015hia} (see also \cite{He:2015yua})\footnote{Recently, inspired by
the development of CHY-approach, a new method to construct all loop integrands for
general massless quantum field theories has been proposed in \cite{Baadsgaard:2015twa}.}.

Although conceptually the CHY approach is remarkable and very useful
for many theoretical studies of properties of  scattering
amplitudes, when applying to real evaluation, one faces the problem
of solving scattering equations, which has $(n-3)!$ solutions in
general. Furthermore, when $n\geq 6$, one encounters polynomials of
degree exceeding five, rendering analytic solutions in radicals
hopeless. Nevertheless, while the solutions can be very complicated,
when putting them back into the CHY integrand and summing up, one
obtains simple rational functions. These observations have led
people to wonder if there is a better way to evaluate the
CHY-integrand {\it without explicitly solving} the scattering
equations. In \cite{Kalousios:2015fya}, using classical formulas of
Vieta, which relate the sums of roots of polynomials to the
coefficients of these polynomials, analytic expression can be
obtained without solving roots explicitly. More general algorithms
are given by two works. In one approach \cite{Cachazo:2015nwa},
using known results for scalar $\phi^3$ theory, one can iteratively
decompose the 4-regular graph determined by the corresponding
CHY-integrand to building blocks related to $\phi^3$ theory, thus
finishing the evaluation. In another approach
\cite{Baadsgaard:2015voa, Baadsgaard:2015ifa}, by careful analysis
of pole structures, the authors wrote down a mapping rule, so that
from the related  CHY-integrand, one can read out contributions of
corresponding Feynman diagrams.

Both approaches are powerful and have avoided the need of solving
the scattering equations explicitly. Furthermore, based on these
perspectives, especially the mapping rule, one can use Feynman
diagrams to construct the CHY-integrand. These results produce a
very interesting phenomenon: {\it two different CHY-integrands can
produce the same result}. For example, there are two very different
CHY-integrands for scalar $\phi^4$ theory: one is given in
\cite{Cachazo:2014xea}, while another one is given in
\cite{Baadsgaard:2015voa, Baadsgaard:2015ifa}. We are naturally led
to wonder how to explain the equivalence of different
CHY-integrands.

In fact, as a rational function of coordinates $z_i$ on a Riemann
surface, the equivalence can occur on three different levels.
\begin{enumerate}
\item
  At the first level, their equivalence is pure algebraic, i.e.,
  through some algebraic manipulations, one rational expression
  can be transformed to another one. For example, for 4-point
  amplitudes of $\phi^3$ theory, on the one hand we have the
  integrand $I_1={1\over z_{12}z_{23} z_{34}z_{41}}{1\over
  z_{12}z_{23} z_{34}z_{41}}$ where we have defined
  $z_{ij}=z_i-z_j$ which gives a contribution of ${1\over
  s_{12}}+{1\over s_{41}}$. On the other hand, we have the
  integrands $I_{21}={1\over z_{12}z_{23} z_{34}z_{41}}{1\over
  z_{12}z_{24}z_{43}z_{31}}$ which gives $-{1\over s_{12}}$ and
  the integrand $I_{22}={1\over z_{12}z_{23}
  z_{34}z_{41}}{1\over  z_{13}z_{32}z_{24} z_{41}}$ which gives
  $-{1\over s_{41}}$. It is easy to check algebraically that
  $I_1=-I_{21}-I_{22}$. Equivalences at this level is of course
  rather trivial and in order to proceed to the other two levels
  of equivalences, we need to change our viewpoint to algebraic
  geometry, i.e., to transform the scattering equations to a set
  of polynomials of $(n-3)$ variables, defining an ideals ${\cal
  I}$;

\item
  The difference of two CHY-integrands can be written as
  $J(I_1-I_2)={P\over Q}$ where both $P,Q$ are polynomials and
  $J$ is the Jacobian we will review shortly. If $P$ belongs to
  the ideal ${\cal I}$, then for each solution of the scattering
  equations $P=0$, thus at the second level we say that  $I_1$
  is equivalent to $I_2$;

\item
  However, in practice, most of the time something more
  complicated happens and we find that though $P$ does not
  belong to the ideal ${\cal I}$ and $J(I_1-I_2)=0$ when and
  only when we sum over all solutions. If this happens, we say
  that  $I_1$ is equivalent to $I_2$ at the third level. It is
  clear that this is the most involved situation, and indeed, in
  practice this is the most frequently encountered.
\end{enumerate}

Motivated by the above considerations and bearing in mind that
indeed the most conducive perspective on studying the scattering
equations is through the language of algebraic varieties and
polynomial ideals \cite{Dolan:2014ega,He:2014wua}, we turn to this
method of attack. The above problem thus translates to finding the
sum over the rational function ${P \over Q}$ evaluated at the roots
of a zero-dimensional ideal $I$, and testing whether the sum is
zero. Luckily, there is a theorem in commutative algebra, due to
Stickelberger, which addresses the situation
\cite{Stickelberger:1989abc}. We will discuss the theorem and the
associated algorithm in illustrative detail.  It turns out that this
method not only checks the equivalence at the third level, but also
evaluates the integration without solving the scattering equations.
In this sense, it is in the spirit of the methods in
\cite{Cachazo:2015nwa} and \cite{Baadsgaard:2015voa,
Baadsgaard:2015ifa}. Although it is sometimes less efficient
compared to these two methods, it does provide a very different
angle to approach the problem and could have very advantageous
repercussions.

The structure of the paper is as follows. We begin with a brief
review of the tree-level scattering equations in \S\ref{s:rev},
before laying down the foundations of the theory of zero-dimensional
ideals in \S\ref{s:math}, especially that of companion matrices. We
then illustrate the technique with ample computational examples in
\S\ref{s:eg}, before concluding with remarks in \S\ref{s:conc}.

\section{Review of Tree-level Scattering Equations}\label{s:rev}
In this section, we offer a brief review of tree-level scattering
equations and the reader is referred to
\cite{Cachazo:2013gna,Cachazo:2013hca, Cachazo:2013iea,
Cachazo:2014xea} for details. The scattering equations are given by
\bea
    {\cal E}_a\equiv \sum_{b\neq a} {s_{ab}\over z_a-z_b}=0,~~~~~a=1,2,...,n \ ,
    \label{SE-def}
\eea where $s_{ab}=(k_a+k_b)^2= 2k_a\cdot k_b$, and $k_a$ with
$a=1,2,...,n$ are $n$ massless momenta for $n$-external particles
and $z_i$ are complex variables living on $\IC\IP^1$ with $n$
punctures. Although there are $n$ equations, only $(n-3)$ of them
are linear independent after using the momentum conservation and
massless conditions which translate to the following three relations
\bea \sum_a {\cal E}_a=0,~~~~\sum_a {\cal E}_a z_a =0,~~~~\sum_a
{\cal E}_a z_a^2=0 \ , \label{SE-rel} \eea which are, in fact, the
consequence of the $SL(2,\IC)$ symmetry on the $\IC\IP^1$. Because
of this, we can insert only $(n-3)$ delta-function. To make sure the
result does not depend on which three equations have been removed,
we make following combination and define the measure\footnote{A nice
explanation of this fact can be found in \cite{Dolan:2013isa,
Dolan:2014ega}} \bea \Omega({\cal E})\equiv z_{ij}z_{jk}z_{ki}
\prod_{a\neq i,j,k}\delta\left( {\cal E}_a\right) \ ,
\label{measure-Omega} \eea with $z_{ij}=z_i-z_j$. With the above,
the general tree-level amplitude is given by
\bea
    {\cal A}_n& = & \int {\prod_{i=1}^n dz_i\over {\rm
vol}(SL(2,\IC))} \Omega({\cal E}) {\cal F}(z)
    =  \int {\prod_{i=1}^n dz_i\over d\omega}  \Omega({\cal E}) {\cal F}(z) \ ,
    \label{gen-A}
\eea where $d\omega={d z_r d z_s d z_t\over z_{rs} z_{st} z_{tr}}$
comes after we use the M\"obius $SL(2,\IC)$ symmetry to fix the
location of three of the variables $z_r, z_s, z_t$ by the
Faddeev-Popov method. Different QFTs give different forms of the
{\it CHY-integrand ${\cal F}(z)$}. Invariance under the M\"obius
transformation requires ${\cal F}(z)$ to have proper transformation
behaviors, i.e., under  $z'={az+b\over cz +d}$, we have \bea
    {\cal F}(z)\to \left(\prod_{i=1}^{n} {(c z_i+d)^4\over (ad-bc)^2} \right){\cal F}(z) \ .
\eea

To simplify expression \eref{gen-A} further, we integrate out the
delta-functions to arrive at the key expression
\bea
    {\cal A}_n=\sum_{sol} {z_{ij} z_{jk}z_{ki} z_{rs} z_{st} z_{tr}\over (-)^{i+j+k+r+s+t} |\Phi|_{ijk}^{rst}}{\cal F} \ ,
    \label{gen-A-2}
    \eea
where three arbitrary indices $i,j,k$ correspond to three removed
scattering equations while three arbitrary indices $r,s,t$
correspond to the above mentioned three fixed locations. The sum is
over the solution set of the scattering equations, which is
generically a discrete set of points. Furthermore, in the above, the
Jacobian matrix $\Phi$ is calculated as ($a$ for rows and $b$ for
column) \bea \Phi_{ab}= {\partial {\cal E}_a\over \partial z_b}=
\left\{
\begin{array}{ll}
  {s_{ab}\over z_{ab}^2} & a\neq b\\
  -\sum\limits_{c\neq a}{s_{ac}\over z_{ac}^2}~~~~&  a=b\end{array}\right. \ ,
  \label{Phi-Jacobi}
\eea
and $|\Phi|_{ijk}^{rst}$ is the determinant of $\Phi$ after removing the $i$-th, $j$-th and $k$-th rows and $r$-th, $s$-th and $t$-th columns.

\paragraph{Specific Examples: }
Now we list some examples in the literatures \cite{Cachazo:2013hca,
Cachazo:2013iea} (more can be found in \cite{Cachazo:2014xea}).
According to the CHY formula, the integrand unifying scalars($b=0$),
gluons($b=1$) and gravitons ($b=2$) is given by
\def\pf{\mbox{Pf~}}
%
%
\bea
\mathcal{F}_{b,n}=\Big(\sum_{\alpha\in S_n / \IZ_n}
{\Tr(T^{\alpha(1)}\ldots T^{\alpha(n)})\over z_{\alpha (1)\alpha
(2)}\ldots
  z_{\alpha(n)\alpha(1)}}\Big)^{2-b}\Big(\pf'~\Psi\Big)^b \ ,
\eea where the sum is over permutations on $n$ elements by the
symmetric group $S_n$, up to cyclic ordering of $\IZ_n$, $\Psi$ is a
$2n\times 2n$ antisymmetric matrix defined by $\Psi= {\small \left(
       \begin{array}{cc}
         A & -C^{t} \\
         C & B \\
       \end{array}
\right)}$ (where $t$ is the transpose of the matrix), with $A,B,C$ being $n\times n$ matrices with components
\bea
A_{ab}=\left\{ \begin{array}{c}
                       {k_a\cdot k_b\over z_a-z_b}  \\
                     0
                    \end{array}
\right. ~~,~~B_{ab}=\left\{ \begin{array}{c}
                       {\epsilon_a\cdot \epsilon_b\over z_a-z_b} \\
                      0~~~~
                    \end{array}
\right.~~,~~C_{ab}=\left\{ \begin{array}{c}
                       {\epsilon_a\cdot k_b\over z_a-z_b}  \\
                      -\sum_{c\neq a}{\epsilon_a\cdot k_c\over z_a-z_c}
                    \end{array}
\right.~~\mbox{for}~~ \begin{array}{c}
                        a\neq b \\
                        a=b
                      \end{array} \ ,
\eea
and $\pf'~\Psi$ is
the reduced Pfaffian (square-root of the determinant) of $\Psi$ defined by
\bea
\pf'~\Psi=2{(-1)^{i+j}\over z_i-z_j}\pf \Psi^{ij}_{ij} \ ,
\eea
where $1\leq i,j\leq n$ and $\Psi^{ij}_{ij}$ is the matrix $\Psi$
removing rows $i,j$ and columns $i,j$.
We recall that the Pfaffian of a $2n\times 2n$ antisymmetric matrix can be computed as
\bea
\pf\Psi={1\over 2^n n!}\sum_{\sigma\in
S_{2n}}\mbox{sgn}(\sigma)\prod_{i=1}^{n}\Psi_{\sigma_{2i-1}\sigma_{2i}}~,
\eea
where $\mbox{sgn}(\sigma)$ is the signature of $\sigma \in S_{2n}$. Importantly, $\pf \Psi^{ij}_{ij}$ is
non-zero on the solutions of scattering equations, while $\pf'~\Psi$
is independent of the choice of $i,j$.

Specifically, we have that
\def\YM{{\tiny\mbox{YM}}}
\def\g{{\tiny \mbox{G}}}
\def\Det{\mbox{Det}}
\begin{itemize}
\item  For color-ordered bi-adjoint
scalar $\phi^3$ theory,
\bea
\mathcal{F}_{\phi^3}={1\over z_{12}^2z_{23}^2\cdots
  z_{n1}^2}~.~~~\label{phi3-integrand}
\eea

\item For color-ordered  Yang-Mills theory with ordering $\{1,2,...,n\}$,
  \bea
  \mathcal{F}_{\YM}={1\over z_{12}z_{23}\cdots z_{n1}}\pf'~\Psi~.
  \eea

\item For gravity,
  \bea
  \mathcal{F}_{\g}=(\pf'~\Psi)^2={4\over
    (z_i-z_j)^2}\Det(\Psi_{ij}^{ij}) \ .
  \eea
\end{itemize}

Having presented the above examples, let us go back to
\eref{gen-A-2}. As is clear from the expression, the right hand side
is a rational function in the complex variables $z_i$. To employ
methods developed in algebraic geometry, we need to associate
solutions to a zero-dimensional algebraic variety defined by some
polynomials. In other words, we should rewrite ${\cal E}_a$ defined
in \eref{SE-def} to an equivalent polynomial system. This has been
done in a beautiful paper \cite{Dolan:2014ega}, where it has been
shown that scattering equations are equivalent to following set of
polynomials
\bea 0= h_m\equiv \sum_{S\in A, |S|=m} k_S^2 z_S~,~~~~2\leq m\leq
n-2~,~~~~ \label{DG-2} \eea
where the sum is over all ${n!\over (n-m)! m!}$ subsets $S$ of
$A=\{1,2,...,n\}$ with exactly $m$ elements and $k_S=\sum_{b\in S}
k_b$ and $z_S=\prod_{b\in S} z_b$. The algebraic geometry, notably
the affine Calabi-Yau properties of \eqref{DG-2}, has been
investigated in \cite{He:2014wua}.

A very useful observation made in \cite{Dolan:2013isa,
Dolan:2014ega} is that {\it If all $k_S^2\neq 0$, then values of
$z_a$ are all  distinct}. The set \eref{DG-2} has not fixed gauge.
One of the choice of gauge will be to set, as is standard with
points on $\IC\IP^1$, the three points $z_1=\infty$, $z_2=1$ and
$z_n=0$. Under this choice, the set of polynomial is reduced to \bea
\W h_{1\leq m\leq n-3} \equiv \lim_{z_1\to \infty} { h_{m+1}\over
z_1}= \sum_{S\in A/\{1,n\}, |S|=m} (k_S+k_1)^2 z_S|_{z_2=1,z_n=0} \
, \label{DG-2-1} \eea In summary, $\tilde{h}$ defines a {\bf
zero-dimensional ideal} in the polynomial ring in $n-3$ variables.
Then, using the standard {\bf B\'ezout's theorem}, the number of
points in this ideal (solutions of the scattering equation) is
$\prod_{m=1}^{n-3} {\rm deg}(\W h_m)=(n-3)!$.


Instead of computing the amplitude with formula (\ref{gen-A-2}) by
summing over all solutions of scattering equations, we will  show in next section that,
using the companion-matrix method, we can compute the amplitude
$\cA_n=\sum_{sol}{P\over Q}$ as the trace of certain matrix composed of
so-called {\bf companion matrices} $T_{z_i}$
\bea
\cA_n=\Tr(P'|_{z_i\to T_{z_i}}Q'^{-1}|_{z_i\to T_{z_i}}) \ ,
\eea
without the explicit solutions of scattering equations.

\section{The Mathematical Framework}\label{s:math}
As mentioned in the introduction, it is expedient to consider the problem within the framework of ideal theory.
Our problem is thus the following.
\begin{quote}
  {\bf Problem: } Let $I = \left< f_i \right>$ be a zero-dimensional ideal in $R = \IC[x_1,\ldots, x_n]$ generated by $f_{i=1,2,\ldots,k}(x_1,\ldots,x_n) \in R$ and let $r(x_1,\ldots,x_n)$ be an arbitrary rational function in the fraction field of $R$.
  Because $\dim_{\IC} I = 0$, $I = \sqcup_{j=1}^N \{z_j \}$ is a discrete set of, say $N$, points. We wish to evaluate
  \[
  \sum\limits_{j=1}^N p(z_j)
  \]
  where each summand is an evaluation of $p$ at one of the discrete set of zeros $z_j$. In particular we wish to test whether this sum is 0.
  This is the level 3 equivalence mentioned in the introduction.
\end{quote}

Of course, the idea is to solve this {\it without} explicitly
finding the roots $z_j$. This can be done using the technique of
companion matrices \cite{sturmfels:2002abc} (cf.~also
\cite{parrilo:2006abc}). Suppose a Gr\"obner basis for $I$ has been
found for some appropriate monomial ordering and $\B$ is an
associated monomial basis for $I$, which can be seen as a vector
space of dimension $d$. Then the multiplication map by the
coordinate variable $x_i$
\begin{eqnarray}
  \nn & R/I & \longrightarrow  R/I \\
T_i : & f & \longrightarrow  x_i f
\end{eqnarray}
is an endomorphism of quotient rings. In the basis $\B$ of
monomials, this is a $d \times d$ matrix and is called a {\bf
companion matrix}. Clearly, $\{T_i\}$ all mutually commute and thus
can be simultaneously diagonalized. We have the following
\cite{Stickelberger:1989abc}:
\begin{theorem}
  [Stickelberger]
  The complex roots $z_i$ of $I$ are the vectors of simultaneous eigenvalues of the companion matrices $T_{i=1,\ldots,n}$, i.e., the corresponding zero dimensional variety consists of the points:
  \[
  \cV(I) = \left\{
  (\lambda_1, \ldots, \lambda_n) \in \IC^n : \exists v \in \IC^n \forall i :
  T_i v = \lambda_i v
  \right\} \ .
  \]
\end{theorem}
We point out that the original statement of the theorem is in terms
of annihilators in algebraic number theory and is perhaps a little
abstruse. Fortunately, the computational algebraic-geometry
community has rephrased this into the readily usable form of
companion matrices \cite{sturmfels:2002abc,M2}. In particular, we
have the following important consequence:
\begin{corollary}
  Our desired quantity
  \[
  \sum\limits_{j=1}^N r(z_j) = \tr [r\left( T_1, \ldots, T_n \right)]
  \]
  where the evaluation of the rational function $r$  on the matrices $T_i$ is without ambiguity since they mutually commute.
\end{corollary}
We remark that because $r$ is rational, whenever the companion matrices appear in the denominator, they are to be understood as the inverse matrix.

\subsection{Warmup}
Before proceeding to examples in our context, we present two simple
exercises to demonstrate our algorithm. Computations can be made in
Macaulay2 \cite{M2} or Singular \cite{DGPS}, or the latter's
interface with Mathematica \cite{Gray:2008zs}. Let
\begin{equation}
  I := \left< xy - z, yz-x, zx - y \right> \subset R = \IC[x,y,z] \ .
\end{equation}
We know, of course, that there are 5 roots
\begin{equation}\label{eg-sol}
  \cV(I) =
  \{
  (0,0,0),
  (-1,-1,1), \
  (-1,1,-1), \
  (1,-1,-1), \
  (1,1,1)
  \} \ .
\end{equation}

Now we consider two functions, where one is polynomial and another,
rational: \bea\label{egsol2} p(x,y,z) = 3x^3y + xyz, ~~~~Q(x,y,z)= {
3x^3y + xyz\over 2 x y^2 + 4 z^2 +1} \ . \eea It is easy to find,
after summing over the solutions, that \bea \sum_{\cV(I)}p=4 \ ,
\qquad \sum_{\cV(I)}Q={20\over 21} \ .\eea We now show how the
companion matrices work without finding the roots \eqref{eg-sol}
explicitly.

In the lex ordering of $x \prec y \prec z$, the Gr\"obner basis and the monomial basis are, respectively,
\begin{equation}
  GB(I) = \left<
  z^3-z, yz^2-y, y^2-z^2, x-yz
  \right> \ ;
  \quad
  \B = \{ 1, y, yz, z, z^2 \} \ .
\end{equation}
Therefore, we have that, in the quotient ring $R/I$,
\begin{equation}
  x . \B = \{yz, z, z^2, y, yz\} \ , \quad
  y . \B = \{y, z^2, z, yz, y \} \ , \quad
  z . \B = \{z, yz, y, z^2, z \} \ ,
\end{equation}
so that
\begin{equation}
  T_x = {\tiny
  \left(
\begin{array}{ccccc}
 0 & 0 & 1 & 0 & 0 \\
 0 & 0 & 0 & 1 & 0 \\
 0 & 0 & 0 & 0 & 1 \\
 0 & 1 & 0 & 0 & 0 \\
 0 & 0 & 1 & 0 & 0 \\
\end{array}
\right)
  }
  \ , \quad
T_y={\tiny
\left(
\begin{array}{ccccc}
 0 & 1 & 0 & 0 & 0 \\
 0 & 0 & 0 & 0 & 1 \\
 0 & 0 & 0 & 1 & 0 \\
 0 & 0 & 1 & 0 & 0 \\
 0 & 1 & 0 & 0 & 0 \\
\end{array}
\right)
} \ , \quad
T_z = {\tiny
\left(
\begin{array}{ccccc}
 0 & 0 & 0 & 1 & 0 \\
 0 & 0 & 1 & 0 & 0 \\
 0 & 1 & 0 & 0 & 0 \\
 0 & 0 & 0 & 0 & 1 \\
 0 & 0 & 0 & 1 & 0 \\
\end{array}
\right) } \ .
\end{equation}
Therefore, the sum over the roots of $p$ is
\begin{equation}
  \tr\left(
  3 T_x^3 T_y + T_x T_y T_z
  \right)
  = 4
\end{equation}
and we have nice agreement with \eqref{egsol2}.

For the $Q$, the numerator is
$N = 3 T_x^3 T_y + T_x T_y T_z = {\scriptsize \left(
\begin{array}{ccccc}
 0 & 0 & 0 & 3 & 1 \\
 0 & 1 & 3 & 0 & 0 \\
 0 & 3 & 1 & 0 & 0 \\
 0 & 0 & 0 & 1 & 3 \\
 0 & 0 & 0 & 3 & 1
\end{array}
\right)}$, while the denominator is
$D = 2 T_x T_y^2 + 4 T_z^2 + I = {\scriptsize \left(
\begin{array}{ccccc}
 1 & 0 & 2 & 0 & 4 \\
 0 & 5 & 0 & 2 & 0 \\
 0 & 0 & 5 & 0 & 2 \\
 0 & 2 & 0 & 5 & 0 \\
 0 & 0 & 2 & 0 & 5
\end{array}
\right)}$.
Thus we calculate $ {\rm Tr}( N D^{-1})= {20\over 21}$, which is the right answer on comparing with \eqref{egsol2}.

Before going to examples of scattering equations, let us give some
remarks. First, the theorem in its original form is for polynomial
test functions $r$, while  functions we will meet in scattering
equations are {\it rational functions}, i.e., the form ${P\over Q}$
with both $P,Q$ are polynomials. Luckily, the theorem and corollary
can be generalized trivially since we can diagonalize companion
matrices simultaneously because the next remark.

Now, there is a second part of the theorem which states that the
companion matrices can be simultaneously diagonalized {\it if and
only if} the ideal $I$ is a radical ideal. That is, there are no
multiple roots. However, as shown in \cite{Dolan:2013isa}, if all
$k_S^2\neq 0$, the solutions of $z_i$ will all be different, so we
indeed have a radical ideal and find simultaneous eigenvalues
readily.

Third, since there are  $(n-3)!$ solutions, the size of $T_i$ will
be in general $d=(n-3)!$ which will become very large with $n$.
Although with this counting,  the efficiency of the method may be
arguable, it does make the following property manifest: {\it after
summing over all solutions, the final result must be rational
functions of $k,\eps$}.

\section{Illustrative Examples}\label{s:eg}

In the following, we will use several examples to demonstrate the
companion matrix method. The $n=4$ case is simple. The companion
matrix is 1-dimensional, equaling to the single solution of
scattering equations. We compute the amplitudes in scalar $\phi^3$,
Yang-Mills and gravity theories to show the validity of the method.
For $n=5$, we first study the amplitude of scalar $\phi^3$ theory,
and show that the amplitude-level identity can be understood by the
fact that the trace of matrix is a linear mapping, and use it the
explain a 7-point identity proposed in \cite{Baadsgaard:2015ifa}.
For the amplitudes of Yang-Mills and gravity theories, we will show
that the companion matrix method indeed produce the correct
amplitudes.

For $n=6$, the scalar $\phi^3$ theory will be shown to detect the
pole structures so that the amplitude can be constructed by setting
appropriate kinematics. The Next-MHV gluon amplitude is also
presented as an example to show the validation of companion matrix
method in a more difficult situation. Finally, for $n=7$ amplitudes
of scalar $\phi^3$ theory, we demonstrate that, when companion
matrices are computed in the diagonal form, the diagonal elements of
the integrand matrix (which we recall to be an $(n-3)!\times (n-3)!$
matrix for $n$-points) have one-to-one mapping to the integrand
computed at the $(n-3)!$ solutions of the scattering equations, so
they are not only equivalent at the amplitude level, but also at the
level of each solution as indicated by Stickelberger's theorem.

\subsection{Four-Point Amplitudes}
The $n=4$ case is trivial. There is only $4-3=1$ variable left, so
the companion matrix is just a complex number. Let us remove three
scattering equations $\mathcal{E}_1$, $\mathcal{E}_2$,
$\mathcal{E}_4$ and gauge-fix three points $z_1=\infty,z_2=1$ and
$z_4=0$. The remaining one scattering equation is \bea
\mathcal{E}_3=\sum_{b\neq 3}{s_{3b}\over z_3-z_b}={s_{13}\over
z_3-z_1}+{s_{23}\over z_3-z_2}+{s_{34}\over
  z_3-z_4}={(s_{23}+s_{34})z_3-s_{34}\over z_3(z_3-1)} \ .
\eea

We can define the ideal $I=\ideal{(s_{23}+s_{34})z_3-s_{34}}$ in
$\mathbb{C}[z_3]$. It is a linear function, so the Gr\"{o}bner basis
and monomial basis are trivially
\bea
\gb(I)=\ideal{(s_{23}+s_{34})z_3-s_{34}}~~~,~~~\B=\{1\}~.~~~\eea
The polynomial reduction of $z_{3}\B=\{z_3\}$ over Gr\"{o}bner basis
of ideal $I$ gives the remainder $\{{s_{34}\over s_{23}+s_{34}}\}$.
Thus in the quotient ring, the companion matrix is given by
\bea\label{n4Tz3}
T_{z_3}\B=\{{s_{34}\over
  s_{23}+s_{34}}\}~~\to~~T_{z_3}={s_{34}\over s_{23}+s_{34}}~.~~~
\eea
We now proceed to the three cases of concern.

\subsubsection{Scalar $\phi^3$ Theory}
For the 4-point amplitude in scalar $\phi^3$ theory, we wish to compute (recall that the three points $z_1,z_2,z_4$ have been gauge fixed)
\bea\label{A4scalar}
\cA_4
=\sum_{sol}{z_{12}^2z_{24}^2z_{41}^2\over
|\Phi|^{124}_{124}}{1\over
  z_{12}^2z_{23}^2z_{34}^2z_{41}^2}
=
-\sum_{z_3 \in sol}{1\over
  s_{34}(z_3-1)^2+s_{23}z_3^2} \equiv
\sum_{z_3 \in sol} {P(z_3)\over Q(z_3)}~,~~~
\eea
where we have used the simplification
\bea |\Phi|^{124}_{124}=\Phi_{33}=-{s_{34}\over z_3^2}-{s_{23}\over
(z_3-1)^2}~~\to~~{1\over|\Phi|^{124}_{124}}=-{z_3^2(z_3-1)^2\over
  z_3^2(s_{12}+s_{23})-2z_3s_{12}+s_{12}}~,~~~
\eea
so that the factor ${1/z_{23}^2z_{34}^2}$ cancels the numerator of
$1/|\Phi|^{124}_{124}$.
We see that the final expression is summed over the (discrete) solution set of the scattering equation which is rather trivial here.
The summand is a rational function in the free variable $z_3$ which we define as $P/Q$; of course, $P=1$ here and $Q$ will be used later.

Finally, using the simple expression for the companion matrix  $T_{z_3}$ from \eqref{n4Tz3}, we have
%
\bea \Tr\Big(\frac{P(T_{z_3})}{Q(T_{z_3})}\Big) = \Tr\Big(-{1\over
T_{z_3}T_{z_3}(s_{12}+s_{23})-2T_{z_3}s_{12}+s_{12}}\Big)=-{s_{23}+s_{34}\over
  s_{23}s_{34}}=-{1\over s_{14}}-{1\over s_{12}}~,~~~
\eea
after some identities between Mandelstam variables have been used.
This is indeed the same answer as the standard known result as given in the introduction.

\subsubsection{Yang-Mills Theory}
For 4-point amplitude in Yang-Mills theory, we want to compute
(under gauge-fixing $z_1=\infty, z_2=1,z_4=0$), \bea
\cA^{\YM}_4=\sum_{sol}{z_{12}^2z_{24}^2z_{41}^2\over
|\Phi|^{124}_{124}}{\pf' \Psi_{8\times 8}\over
  z_{12}z_{23}z_{34}z_{41} }\equiv
\sum_{z_3 \in sol}
{P^{\YM}(z_3)\over
  Q^{\YM}(z_3)}~.~~~
\eea
To avoid the
divergence when taking the limit $z_1 \to \infty$, one of the removed
rows(columns) in $\Psi$ should be 1, otherwise some terms in $\pf'
\Psi_{8\times 8}$ would lead to infinity. Let us then choose the
reduced Pfaffian as
\bea \pf' \Psi_{8\times 8 }={-2\over z_1-z_2}\pf
\Psi^{12}_{12}~.~~~\eea
The large $z_1$ dependence of $\pf' \Psi_{8\times 8}$ is then
$1/z_1^2$, and together with the factor from the scalar part, we obtain
a finite integrand when taking the $z_1 \to \infty$ limit.
Explicitly, the
new matrix $\widetilde{\Psi}\equiv \Psi^{12}_{12}$ is a $6\times 6$
matrix,
\bea \widetilde{\Psi}=\left(\begin{array}{cccccc}
 0 & \frac{k_3 k_4}{{z_3}-{z_4}} & -\frac{\epsilon_1 k_3}{{z_1}-{z_3}} &~~~ -\frac{\epsilon_2 k_3}{{z_2}-{z_3}} ~~~& \sum_{c\neq 3}{\epsilon_3k_c\over z_3-z_c} & -\frac{\epsilon_4 k_3}{{z_4}-{z_3}} \\
 \frac{k_3 k_4}{{z_4}-{z_3}} & 0 & -\frac{\epsilon_1 k_4}{{z_1}-{z_4}} & -\frac{\epsilon_2 k_4}{{z_2}-{z_4}} & -\frac{\epsilon_3 k_4}{{z_3}-{z_4}} &\sum_{c\neq 4}{\epsilon_4k_c\over z_4-z_c} \\
 \frac{\epsilon_1 k_3}{{z_1}-{z_3}} & \frac{\epsilon_1 k_4}{{z_1}-{z_4}} & 0 & \frac{\epsilon_1 \epsilon_2}{{z_1}-{z_2}} & \frac{\epsilon_1 \epsilon_3}{{z_1}-{z_3}} & \frac{\epsilon_1 \epsilon_4}{{z_1}-{z_4}} \\
 \frac{\epsilon_2 k_3}{{z_2}-{z_3}} & \frac{\epsilon_2 k_4}{{z_2}-{z_4}} & \frac{\epsilon_1 \epsilon_2}{{z_2}-{z_1}} & 0 & \frac{\epsilon_2 \epsilon_3}{{z_2}-{z_3}} & \frac{\epsilon_2 \epsilon_4}{{z_2}-{z_4}} \\
-\sum_{c\neq 3}{\epsilon_3k_c\over z_3-z_c} & \frac{\epsilon_3 k_4}{{z_3}-{z_4}} & \frac{\epsilon_1 \epsilon_3}{{z_3}-{z_1}} & \frac{\epsilon_2 \epsilon_3}{{z_3}-{z_2}} & 0 & \frac{\epsilon_3 \epsilon_4}{{z_3}-{z_4}} \\
 \frac{\epsilon_4 k_3}{{z_4}-{z_3}} & -\sum_{c\neq 4}{\epsilon_4k_c\over z_4-z_c} & \frac{\epsilon_1 \epsilon_4}{{z_4}-{z_1}} & \frac{\epsilon_2 \epsilon_4}{{z_4}-{z_2}} & \frac{\epsilon_3 \epsilon_4}{{z_4}-{z_3}} & 0 \\
\end{array}\right)~,~~~\eea
whose Pfaffian is given by
\bea \pf \Psi^{12}_{12}&=&\widetilde{\Psi}_{1 6} \widetilde{\Psi}_{2
5} \widetilde{\Psi}_{3 4} - \widetilde{\Psi}_{1 5}
\widetilde{\Psi}_{2 6} \widetilde{\Psi}_{3 4} -
 \widetilde{\Psi}_{1 6} \widetilde{\Psi}_{2 4} \widetilde{\Psi}_{3 5} + \widetilde{\Psi}_{1 4} \widetilde{\Psi}_{2 6} \widetilde{\Psi}_{3 5} +
 \widetilde{\Psi}_{1 5} \widetilde{\Psi}_{2 4} \widetilde{\Psi}_{3 6}\nonumber\\
  &&- \widetilde{\Psi}_{1 4} \widetilde{\Psi}_{2 5} \widetilde{\Psi}_{3 6} +
 \widetilde{\Psi}_{1 6} \widetilde{\Psi}_{2 3} \widetilde{\Psi}_{4 5} - \widetilde{\Psi}_{1 3} \widetilde{\Psi}_{2 6} \widetilde{\Psi}_{4 5} +
 \widetilde{\Psi}_{1 2} \widetilde{\Psi}_{3 6} \widetilde{\Psi}_{4 5} - \widetilde{\Psi}_{1 5} \widetilde{\Psi}_{2 3} \widetilde{\Psi}_{4 6} \nonumber\\
 &&+
 \widetilde{\Psi}_{1 3} \widetilde{\Psi}_{2 5} \widetilde{\Psi}_{4 6} - \widetilde{\Psi}_{1 2} \widetilde{\Psi}_{3 5} \widetilde{\Psi}_{4 6} +
 \widetilde{\Psi}_{1 4} \widetilde{\Psi}_{2 3} \widetilde{\Psi}_{5 6} - \widetilde{\Psi}_{1 3} \widetilde{\Psi}_{2 4} \widetilde{\Psi}_{5 6} +
 \widetilde{\Psi}_{1 2} \widetilde{\Psi}_{3 4} \widetilde{\Psi}_{5 6}
~.~~~\eea
The reduced Pfaffian $\pf'\Psi_{8\times 8}$ in this case is a
rational function with denominator $z_3^2(z_3-1)$. Together with the
factor $1/z_{23}z_{34}=1/z_3(z_3-1)$, they cancel the numerator of
$1/|\Phi|^{124}_{124}$, leaving a $z_3$ in the denominator of
integrand.

Therefore, it is immediate that the numerator of the
integrand comes entirely from the numerator of the reduced Pfaffian:
\bea\label{PYMn4}
P^{\YM}&=&z_3^2\big(-s_{12} \widetilde{\epsilon}_{1, 3}
\widetilde{\epsilon}_{2, 4} - 2 \widetilde{\epsilon}_{3, 4}
\kappa_{1, 3} \kappa_{2, 4} +
 2 \widetilde{\epsilon}_{2, 4} \kappa_{1, 4} \kappa_{3, 2} - 2 \widetilde{\epsilon}_{1, 4} \kappa_{2, 4} \kappa_{3, 2} +
 2 \widetilde{\epsilon}_{2, 4} \kappa_{1, 3} \kappa_{3, 4}\nonumber\\
 &&~~~ + 2 \widetilde{\epsilon}_{2, 4} \kappa_{1, 4} \kappa_{3, 4} -
 2 \widetilde{\epsilon}_{1, 4} \kappa_{2, 4} \kappa_{3, 4} + 2 \widetilde{\epsilon}_{2, 3} \kappa_{1, 3} \kappa_{4, 2} -
 2 \widetilde{\epsilon}_{1, 3} \kappa_{2, 3} \kappa_{4, 2} + 2 \widetilde{\epsilon}_{1, 2} \kappa_{3, 2} \kappa_{4, 2} \nonumber\\
 &&~~~+
 2 \widetilde{\epsilon}_{1, 2} \kappa_{3, 4} \kappa_{4, 2} + 2 \widetilde{\epsilon}_{1, 3} \kappa_{2, 4} \kappa_{4, 3}
\big)+z_3\big(-s_{12} \widetilde{\epsilon}_{1, 4}
\widetilde{\epsilon}_{2, 3} + s_{12} \widetilde{\epsilon}_{1, 3}
\widetilde{\epsilon}_{2, 4} +
 s_{12} \widetilde{\epsilon}_{1, 2} \widetilde{\epsilon}_{3, 4}\nonumber\\
 &&~~~ - 2 \widetilde{\epsilon}_{3, 4} \kappa_{1, 4} \kappa_{2, 3} +
 2 \widetilde{\epsilon}_{3, 4} \kappa_{1, 3} \kappa_{2, 4} - 2 \widetilde{\epsilon}_{2, 4} \kappa_{1, 3} \kappa_{3, 4} -
 2 \widetilde{\epsilon}_{2, 4} \kappa_{1, 4} \kappa_{3, 4} + 2 \widetilde{\epsilon}_{1, 4} \kappa_{2, 3} \kappa_{3, 4} \nonumber\\
 &&~~~+
 2 \widetilde{\epsilon}_{1, 4} \kappa_{2, 4} \kappa_{3, 4} - 2 \widetilde{\epsilon}_{1, 2} \kappa_{3, 4} \kappa_{4, 2} +
 2 \widetilde{\epsilon}_{2, 3} \kappa_{1, 3} \kappa_{4, 3} + 2 \widetilde{\epsilon}_{2, 3} \kappa_{1, 4} \kappa_{4, 3} -
 2 \widetilde{\epsilon}_{1, 3} \kappa_{2, 3} \kappa_{4, 3}\nonumber\\
 &&~~~ - 2 \widetilde{\epsilon}_{1, 3} \kappa_{2, 4} \kappa_{4, 3} +
 2 \widetilde{\epsilon}_{1, 2} \kappa_{3, 2} \kappa_{4, 3}\big)
 -s_{12} \widetilde{\epsilon}_{1, 2} \widetilde{\epsilon}_{3, 4}~,~~~\eea
where $\widetilde{\epsilon}_{i,j}\equiv \epsilon_i\epsilon_j$,
$\kappa_{i,j}\equiv \epsilon_ik_j$.
The denominator of the integrand, on the other hand, is
\bea\label{QYMn4}
Q^{\YM}=z_3^3(s_{12}+s_{23})-2z_3^2s_{12}+z_3s_{12}=z_3Q~,~~~\eea
where $Q$ is the denominator of integrand for the scalar $\phi^3$
theory from \eqref{A4scalar}.

In summary, by computing $\Tr(PQ^{-1}|_{z_3\to T_{z_3}})$, we arrive at
\bea
\cA_{4}^{\YM}&=&\widetilde{\epsilon}_{1, 3}
\widetilde{\epsilon}_{2, 4}-\widetilde{\epsilon}_{1, 4}
\widetilde{\epsilon}_{2, 3}-\widetilde{\epsilon}_{1, 2}
\widetilde{\epsilon}_{3, 4}-{s_{12}\over s_{23}}
\widetilde{\epsilon}_{1, 4} \widetilde{\epsilon}_{2, 3}   -
{s_{23}\over s_{12}} \widetilde{\epsilon}_{1, 2}
\widetilde{\epsilon}_{3, 4}\nonumber\\
&&+{1\over s_{12}}\Big(-2 \widetilde{\epsilon}_{3, 4} \kappa_{1, 4}
\kappa_{2, 3} + 2 \widetilde{\epsilon}_{3, 4} \kappa_{1, 3}
\kappa_{2, 4} -
 2 \widetilde{\epsilon}_{2, 4} \kappa_{1, 3} \kappa_{3, 4} - 2 \widetilde{\epsilon}_{2, 4} \kappa_{1, 4} \kappa_{3, 4} +
 2 \widetilde{\epsilon}_{1, 4} \kappa_{2, 3} \kappa_{3, 4}\nonumber\\
 &&~~~~~~~~~ + 2 \widetilde{\epsilon}_{1, 4} \kappa_{2, 4} \kappa_{3, 4} -
 2 \widetilde{\epsilon}_{1, 2} \kappa_{3, 4} \kappa_{4, 2} + 2 \widetilde{\epsilon}_{2, 3} \kappa_{1, 3} \kappa_{4, 3} +
 2 \widetilde{\epsilon}_{2, 3} \kappa_{1, 4} \kappa_{4, 3} - 2 \widetilde{\epsilon}_{1, 3} \kappa_{2, 3} \kappa_{4, 3} \nonumber\\
 &&~~~~~~~~~-
 2 \widetilde{\epsilon}_{1, 3} \kappa_{2, 4} \kappa_{4, 3} + 2 \widetilde{\epsilon}_{1, 2} \kappa_{3, 2} \kappa_{4,
 3}\Big)\nonumber\\
 &&+{1\over s_{23}}\Big( -2 \widetilde{\epsilon}_{3, 4} \kappa_{1, 4} \kappa_{2, 3} + 2 \widetilde{\epsilon}_{2, 4} \kappa_{1, 4} \kappa_{3, 2} -
 2 \widetilde{\epsilon}_{1, 4} \kappa_{2, 4} \kappa_{3, 2} + 2 \widetilde{\epsilon}_{1, 4} \kappa_{2, 3} \kappa_{3, 4} +
 2 \widetilde{\epsilon}_{2, 3} \kappa_{1, 3} \kappa_{4, 2} \nonumber\\
 &&~~~~~~~~~- 2 \widetilde{\epsilon}_{1, 3} \kappa_{2, 3} \kappa_{4, 2} +
 2 \widetilde{\epsilon}_{1, 2} \kappa_{3, 2} \kappa_{4, 2} + 2 \widetilde{\epsilon}_{2, 3} \kappa_{1, 3} \kappa_{4, 3} +
 2 \widetilde{\epsilon}_{2, 3} \kappa_{1, 4} \kappa_{4, 3} - 2 \widetilde{\epsilon}_{1, 3} \kappa_{2, 3} \kappa_{4, 3} \nonumber\\
 &&~~~~~~~~~+
 2 \widetilde{\epsilon}_{1, 2} \kappa_{3, 2} \kappa_{4, 3}\Big)~.~~~\eea
The pole structures are similar to the scalar $\phi^3$ theory, while
the terms without poles come from the gluon four-vertex.
Of course, by
momentum conservation and the property $\epsilon_ik_i=0$, we can
further write the above result as a function of all independent
kinematics, for example by using identities
$\epsilon_{j}k_4=-\epsilon_{j}k_3-\epsilon_{j}k_2-\epsilon_{j}k_1$
and $\epsilon_{4}k_4=0$. This result agrees with the one computed
directly by Feynman diagrams.

\subsubsection{Gravity}
For the 4-point amplitude in gravity, we want to compute
\bea \cA^{\g}_4=\sum_{sol}{z_{12}^2z_{24}^2z_{41}^2\over
|\Phi|^{124}_{124}}\Det'(\Psi_{8\times
8})=\sum_{sol}{z_{12}^2z_{24}^2z_{41}^2\over
  |\Phi|^{124}_{124}}(\pf' \Psi_{8\times 8})^2
\equiv
\sum_{z_3 \in sol}
{P^{\g}(z_3)\over
Q^{\g}(z_3)}~,~~~\eea
under the gauge-fixing $z_1=\infty, z_2=1,z_4=0$. As in Yang-Mills
theory, we choose the reduced Pfaffian as
\bea \pf' \Psi_{8\times 8 }={-2\over z_1-z_2}\pf
\Psi^{12}_{12}~,~~~
\eea
and as above, we know that the squared reduced Pfaffian $(\pf'\Psi_{8\times 8})^2$ is a rational function with denominator $z_3^4(z_3-1)^2$.
This cancels the numerator of
$1/|\Phi|^{124}_{124}$, leaving a $z_3^2$ in the denominator of
integrand, so that the numerator of integrand equals to the square of
numerator of reduced Pfaffian:
\bea P^{\g}=(P^{\YM})^2~,~~~\eea
while the denominator of integrand is
\bea
Q^{\g}=z_3^4(s_{12}+s_{23})-2z_3^3s_{12}+z_3^2s_{12}=z_3^2Q=z_3Q^{\YM}~.~~~
\eea

Combining all together, we have
\bea {P^{\g}\over Q^{\g}}={(P^{\YM})^2\over
z_3Q^{\YM}}={Q^{\YM}\over z_3}{(P^{\YM})^2\over
(Q^{\YM})^2}=Q{(P^{\YM})^2\over (Q^{\YM})^2}~,~~~\eea
where $Q$ is the denominator of integrand for scalar $\phi^3$
theory from \eqref{A4scalar} and the expressions for $P^{\YM}$ and $Q^{\YM}$ are given in \eqref{PYMn4} and \eqref{QYMn4}.
In the present case of $n=4$, there is only one solution for
scattering equations, and the companion matrix is really
1-dimensional in \eqref{n4Tz3}, so although in general $\Tr(M_1M_2)\neq
\Tr(M_1)\Tr(M_2)$, here we simply have
\bea \Tr\Big({P^{\g}\over Q^{\g}}\Big)=\Tr(Q)\Tr\Big({P^{\YM}\over
Q^{\YM}}\Big)^2=-{s_{12}s_{23}\over
  s_{12}+s_{23}}(\cA^{\YM})^2~.~~~
\eea
By BCJ relation \cite{Bern:2008qj}, we can rewrite this to the familiar one
\bea
\cA^{\g}_4=s_{12}\cA^{\YM}_4(1,2,3,4)\cA^{\YM}_4(1,2,4,3)~,~~~\eea
in agreement with the known result by KLT relation \cite{Kawai:1985xq, Bern:1998sv, BjerrumBohr:2010ta, BjerrumBohr:2010zb, BjerrumBohr:2010yc}.


\subsection{Five-Point Amplitudes}

For $n=5$ amplitudes, there are five scattering equations, but only
two of them are independent. Under the gauge-fixing $z_{1}=\infty$,
$z_2=1$, $z_5=0$, the Dolan-Goddard's formula \cite{Dolan:2013isa}
gives:
\bea
f_1=s_{12}+s_{13}
z_3+s_{14}z_4~~,~~f_2=s_{45}z_3+s_{35}z_4+s_{25}z_3z_4~.~~~
\eea
We can solve these two equations to get two solutions:
\bea
sol_1:~~z_3={-s_{12}s_{25}-s_{13}s_{35}+s_{14}s_{45}-\sqrt{\Delta}\over
2s_{13}s_{25}}~~,~~z_4={-s_{12}s_{25}+s_{13}s_{35}-s_{14}s_{45}+\sqrt{\Delta}\over
2s_{14}s_{25}}~,~~~\nonumber\eea
and
\bea sol_2:~~
z_3={-s_{12}s_{25}-s_{13}s_{35}+s_{14}s_{45}+\sqrt{\Delta}\over
2s_{13}s_{25}}~~,~~z_4={-s_{12}s_{25}+s_{13}s_{35}-s_{14}s_{45}-\sqrt{\Delta}\over
2s_{14}s_{25}}~,~~~\nonumber\eea
where
$\Delta=(s_{12}s_{25}+s_{13}s_{35}-s_{14}s_{45})^2-4s_{12}s_{13}s_{25}s_{35}$.
We can see that, in general the solutions are not rational
functions, as is to be expected from high degree polynomials, though
of course the final result of the sum over these points will be. One
can see that the cancelations and simplifications will be very
involved.

Let us turn to our companion matrix method. Define ideal
$I=\ideal{f_1,f_2}$ in the polynomial ring $\mathbb{C}[z_3,z_4]$, the
Gr\"obner basis of ideal $I$ in {\sl Lexicographic} order $z_3\prec
z_4$ is given by
\bea \gb(I)&=&\ideal{s_{12} s_{45} + s_{12} s_{25} z_4 - s_{13}
s_{35} z_4 + s_{14} s_{45} z_4 + s_{14} s_{25} z_4^2,\nonumber\\
&& ~~~~~~ s_{12} + s_{13} z_3 + s_{14} z_4~~,~~ s_{45} z_3 + s_{35}
z_4 + s_{25} z_3 z_4}~.~~~\eea
The monomial basis in this Gr\"obner basis is $\B=\{1,z_4\}$.
Polynomial reduction of $z_3 \B$ and $z_4 \B$ over $\gb(I)$ gives
the companion matrices $T_{z_3}\B=z_3 \B$, $T_{z_4}\B=z_4\B$ as
\bea T_{z_3}=\left(
               \begin{array}{cc}
                 -{s_{12}\over s_{13}} & -{s_{14}\over s_{13}} \\
                 {s_{12}s_{45}\over s_{13}s_{25}} & {s_{14}s_{45}-s_{13}s_{35}\over s_{13}s_{25}} \\
               \end{array}
             \right)~~~,~~~T_{z_4}=\left(
               \begin{array}{cc}
                 0 & 1 \\
                 -{s_{12}s_{45}\over s_{14}s_{25}} & {s_{13}s_{35}-s_{14}s_{45}-s_{12}s_{25}\over s_{14}s_{25}} \\
               \end{array}
             \right)~,
\eea
which are $2\times 2$ matrices, in accordance with the number of
solutions of scattering equations.

We note that the companion matrices actually formally ``live'' in the ideal $I$ itself by satisfying scattering equations, i.,e.,
\bea f_1&\to& s_{12}I_{2\times
2}+s_{13}T_{z_3}+s_{14}T_{z_4}\nonumber\\
&&=\left(
\begin{array}{cc}
  s_{12} & 0 \\
  0  & s_{12} \\
\end{array}
\right)+\left(
\begin{array}{cc}
  -s_{12} & -s_{14} \\
  {s_{12}s_{45}\over s_{25}} & {s_{14}s_{45}-s_{13}s_{35}\over s_{25}} \\
\end{array}
\right)+\left(
\begin{array}{cc}
  0 & s_{14} \\
  -{s_{12}s_{45}\over s_{25}} & {s_{13}s_{35}-s_{14}s_{45}-s_{12}s_{25}\over s_{25}} \\
\end{array}
\right) = 0_{2\times 2} \ ,
 \eea
and likewise,
$s_{45}T_{z_3}+s_{35}T_{z_4}+s_{25}T_{z_3}T_{z_4}=0_{2\times 2}$.
This is, of course, a general property by construction since the
companion matrices are constructed as multiplication (on a
particular basis), so that substituting into the defining
polynomials would vanish in the quotient ring. The situation is very
much analogous to the classical result of Cayley-Hamilton that a
matrix satisfies its own characteristic polynomial. It is worth to
emphasize this discussion as
\begin{corollary}
The companion matrices satisfy the defining polynomials of the given ideal.
\end{corollary}

The above corollary shows some kind of equivalence between solutions
of scattering equations and companion matrices of monomial basis
over the Gr\"obner basis of scattering equations. With these
companion matrices, we now proceed to compute the trace of the
integrands to obtain the amplitude for different theories.

\subsubsection{Scalar $\phi^3$ Theory}

The 5-point amplitude of scalar $\phi^3$ theory is given by
\bea
\cA_5&=&\sum_{sol}{z_{12}^2z_{25}^2z_{51}^2\over
|\Phi|^{125}_{125}}{1\over
z_{12}^2z_{23}^2z_{34}^2z_{45}^2z_{51}^2}=\sum_{sol}{1\over
  |\Phi|^{125}_{125}(z_3-1)^2(z_3-z_4)^2z_4^2}
\equiv
\sum_{z_3,z_4 \in sol}
    {P(z_3,z_4)\over
      Q(z_3,z_4)}~,~~~
    \eea
where we have used that
\bea |\Phi|^{125}_{125}=\left(-{s_{23}\over (z_3-1)^2}-{s_{34}\over
(z_3-z_4)^2}-{s_{35}\over z_3^2}\right)\left(-{s_{24}\over
(z_4-1)^2}-{s_{34}\over (z_3-z_4)^2}-{s_{45}\over
z_4^2}\right)-{s_{34}^2\over(z_3-z_4)^4}~~~~\nonumber\eea
and as above, defined the appropriate $P$ and $Q$, which are,
explicitly,
\bea P&=&z_3^2(z_4-1)^2~,~~~\nonumber\\
Q&=&\Big(s_{35} ( z_3-1)^2 + s_{23} z_3^2\Big) (z_3 - z_4)^2
\Big(s_{45} ( z_4-1)^2 +
     s_{24} z_4^2\Big)\nonumber\\
     && +
  s_{34} \Big[s_{45} ( z_3-1)^2 z_3^2 (z_4-1)^2\nonumber\\
        && +  z_4^2\Big(z_3^2 \big(s_{24} (z_3-1)^2 + s_{23} (z_4-1)^2\big) +
    s_{35} ( z_3-1)^2 (z_4-1)^2\Big)\Big]~.~~~
  \eea

Now, we wish to compute the trace of the matrix $PQ^{-1}$ upon
substituting $z_3$ and $z_4$ by their associated companion matrices,
instead of summing over all the complicated solutions of the
scattering equations. In other words, we should replace the
variables $z_3,z_4$ as $T_{z_3}, T_{z_4}$ in the integrand, i.e.,
$P'=P|_{z_3\to T_{z_3},z_4\to T_{z_4}}$, $Q'=Q|_{z_3\to
T_{z_3},z_4\to T_{z_4}}$(Hereafter we will always use $P',Q'$ to
denote the matrices after replacing $z_i$ to $T_{z_i}$). The product
of variables $z_3,z_4$ changes to the product of matrices
$T_{z_3},T_{z_4}$, and since the companion matrices are commutable,
their order does not matter in here. Then we should compute the
inverse of matrix $Q'$, and the final result is given by
$\Tr(P'Q'^{-1})$.

Recalling that the physical poles appearing in the color-ordered
amplitude are $s_{12},s_{23},s_{34},s_{45},s_{15}$, we can define
them as the independent Mandelstam variables, and rewrite all the
other Mandelstam variables in $P,Q,T_{z_3},T_{z_4}$ by using
following identities:
%
%
%
\bea
&&s_{35}=s_{12}-s_{34}-s_{45}~~,~~s_{24}=s_{15}-s_{23}-s_{34}~~,~~s_{25}=s_{34}-s_{15}-s_{12}~,~~~\nonumber\\
&&s_{14}=s_{23}-s_{45}-s_{15}~~,~~s_{13}=s_{45}-s_{12}-s_{23}~.~~~\eea
%
After some algebraic manipulation, readily performed by Mathematica,
we obtain
\bea \Tr(P'(T_{z_3},T_{z_4})Q'^{-1}(T_{z_3},T_{z_4}))={1\over
s_{15}s_{23}}+{1\over s_{12}s_{34}}+{1\over s_{15}s_{34}}+{1\over
s_{12}s_{45}}+{1\over s_{23}s_{45}}~,~~~\eea
which agrees with the known result \cite{Baadsgaard:2015voa, Baadsgaard:2015ifa}.

\begin{figure}
\centering
  \includegraphics[width=2in]{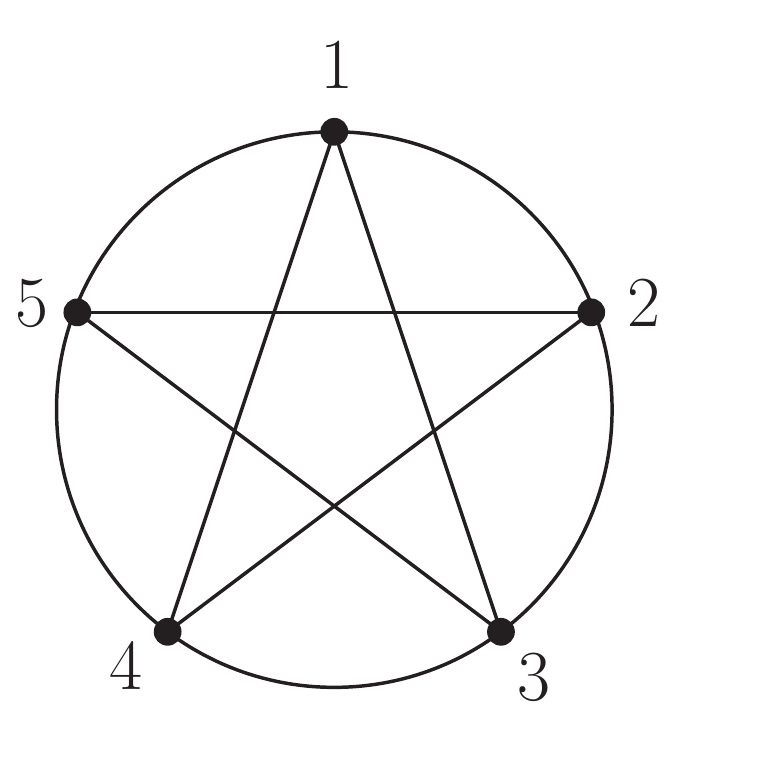}\\
  \caption{The pentacle diagram representing the CHY-integrand defined
  by the two-cycles \{(1,2,3,4,5),(1,3,5,2,4)\}.}\label{pentacle}
\end{figure}

Let us further consider an example, corresponding to the
two-cycles\footnote{Each cycle defines an expression, e.g.,
$\mbox{Cycle}_a(1,3,5,2,4)={1/( z_{13} z_{35} z_{52} z_{24}
z_{41})}$, and the {\sl two-cycles} denotes the expression given by
$\mbox{Cycle}_a\mbox{Cycle}_b$. }
$\{(1,2,3,4,5),(1,3,5,2,4)\}$, in the language of \cite{Cachazo:2015nwa, Baadsgaard:2015voa, Baadsgaard:2015ifa}. 
%
Using the CHY-integrand defined by above two-cycles, we have
\bea \cA'_5&=&\sum_{sol}{z_{12}^2z_{25}^2z_{51}^2\over
|\Phi|^{125}_{125}}{1\over z_{12}z_{23}z_{34}z_{45}z_{51}}{1\over
z_{13}z_{35}z_{52}z_{24}z_{41}}\equiv{P_1(z_3,z_4)\over
Q_1(z_3,z_4)}~,~~~\eea
which is represented by the so-called pentacle diagram (shown in Figure
\ref{pentacle})
from the view of integration rules.
Using the mapping rule
given in \cite{Baadsgaard:2015voa, Baadsgaard:2015ifa}, the answer
is known to be zero. By directly computing the trace, we indeed find
that $\Tr(P'_1Q'^{-1}_1)=0$ and confirms this result.

In fact, for this example, although CHY-integrands of $\cA_5$ and $\cA'_5$ are different, after simplification, their difference appears only in the numerator, i.e.,
\bea P_1=z_3z_4(1-z_3)(1-z_4)(z_3-z_4)~~,~~Q_1=Q~.~~~\eea
Since the trace of matrix is a linear mapping, in particular
$\Tr(M_1+M_2)=\Tr(M_1)+\Tr(M_2)$, relations between results of
different integrands should also have hints in the integrand level.
For example, let us consider the following three CHY-integrands
defined by three two-cycles $\a_2\equiv
\{(1,2,3,4,5),(1,2,3,5,4)\}$, $\a_3\equiv
\{(1,2,3,4,5),(1,2,4,5,3)\}$ and $\a_4\equiv
\{(1,2,3,4,5),(1,3,2,5,4)\}$. With some calculations, we find
\bea {\cal A}_5 (\a_2) & = &¡¡\sum_{sol}{z_{12}^2z_{25}^2z_{51}^2\over
|\Phi|^{125}_{125}}{1\over z_{12}z_{23}z_{34}z_{45}z_{51}}{1\over z_{12}z_{23}z_{35}z_{54}z_{41}}\equiv\sum_{sol} {P_2\over Q}, \nonumber \\
\cA_5(\a_3) & = &\sum_{sol}{z_{12}^2z_{25}^2z_{51}^2\over |\Phi|^{125}_{125}}{1\over z_{12}z_{23}z_{34}z_{45}z_{51}}{1\over z_{12}z_{24}z_{45}z_{53}z_{31}}\equiv \sum_{sol}{P_3\over Q},\nn \\
\cA_5(\a_4) & = &  \sum_{sol}{z_{12}^2z_{25}^2z_{51}^2\over
|\Phi|^{125}_{125}}{1\over
  z_{12}z_{23}z_{34}z_{45}z_{51}}{1\over z_{13}z_{32}z_{25}z_{54}z_{41}}\equiv
     \sum_{sol} {P_4\over Q}      \eea
where they share the same denominator $Q$, but different numerators
\bea P_2=z_3 (z_3 - z_4) (z_4-1)^2~~,~~P_3= z_3 (z_3-1)(z_3 - z_4)
(z_4-1)~~,~~P_4=z_3^2 (z_3 - z_4) (z_4-1)^2~.~~~\eea
After putting back the companion matrices, we find that
\bea &&{\cal A}_5 (\a_2)=
 \Tr(P'_2Q'^{-1})={1\over s_{12}s_{45}}+{1\over
s_{23}s_{45}}~,~~~\nonumber\\
&&{\cal A}_5 (\a_3)= \Tr(P'_3Q'^{-1})={1\over s_{12}s_{45}}~,~~{\cal
A}_5 (\a_4)= \Tr(P'_4Q'^{-1})={1\over
s_{23}s_{45}}~.~~~\nonumber\eea
Realizing that the polynomials have the simple relation
\bea P_3+P_4-P_2=P_1~,~~~\eea
we obtain the identity amongst these amplitudes as
\bea
&&\Tr(P'_2Q'^{-1})=\Tr((P'_3+P'_4-P'_1)Q'^{-1})=\Tr(P'_3Q'^{-1})+\Tr(P'_4Q'^{-1})+\Tr(P'_1Q'^{-1})\nonumber\\
&&\to
\cA_{5}(\alpha_2)=\cA_{5}(\alpha_3)+\cA_{5}(\alpha_4)+0~.~~~\eea
Above example demonstrates an idea how to find relations among
different amplitudes. Starting from different CHY-integrands, we can
equalize their denominators by multiplying proper polynomial both at
the denominator and the numerator. After that, the relations among
different amplitudes can be understood from the relations among
different numerators.


Let us demonstrate above idea by another example, i.e.,  the 7-point
amplitude-level identity given by eq.(3.7) of
\cite{Baadsgaard:2015ifa}, viz., amplitude obtained from the
CHY-integrand
\bea {1\over z_{12}z_{23}z_{34}z_{45}z_{56}z_{67}z_{71}}{1\over
z_{12}z_{27}z_{74}z_{46}z_{65}z_{53}z_{31}}~~~~\label{twocycle1}\eea
is identical to the sum of following two amplitudes obtained from two CHY-integrand
%
\bea {1\over z_{12}z_{23}z_{34}z_{45}z_{56}z_{67}z_{71}}{1\over
z_{12}z_{56}z_{37}z_{46}}\Big({1\over z_{14}z_{27}z_{35}}+{1\over
  z_{25}z_{74}z_{31}}\Big)~.~~~\label{twocycle2}
\eea
Under gauge-fixing $z_1=\infty$, $z_2=1$, $z_7=0$ and excluding the
1-st, 2-nd and 7-th scattering equations, the Jacobian is
\bea
    {1\over |\Phi|^{127}_{127}}={\prod_{i=3}^{6}z_i^2( z_i-1)^2
\prod_{3\leq i<j\leq 6}(z_i - z_j)^2\over
   Q}~.~~~\eea
Thus we can immediately get the numerator of integrand after
inserting the above three terms. The first term gives
\bea &&P_1=z_3z_5(z_4-1)(z_5-1)(z_6-1)(z_3-z_6)\prod_{i=3}^{6}z_i(
z_i-1) \prod_{3\leq i<j\leq 6}^{i\neq 5}(z_i - z_j)~,~~~\eea
while the other two terms give
\bea &&P_2= z_4z_5(z_4-1)(z_5-1)(z_6-1)(z_3-z_6)\prod_{i=3}^{6}z_i(
z_i-1) \prod_{3\leq i<j\leq 6}^{i\neq 5}(z_i - z_j)~,~~~\\
&&P_3=-z_5(z_4-1)(z_6-1)(z_3-z_5)(z_3-z_6)\prod_{i=3}^{6}z_i( z_i-1)
\prod_{3\leq i<j\leq 6}^{i\neq 5}(z_i - z_j)~.~~~\eea
Note that
\bea &&P_1-P_2-P_3\\
 &=&z_5(z_4-1)(z_6-1)(z_3 - z_6)(z_4 - z_5 + z_3
z_5 - z_4 z_5) \prod_{i=3}^{6}z_i( z_i-1) \prod_{3\leq i<j\leq
6}^{i\neq 5}(z_i - z_j)~,~~~\nonumber\eea
while the trace  $\Tr((P_1'-P'_2-P'_3)Q'^{-1})$ is zero. Note also
the following decomposition
\bea z_4-z_5+z_3z_5-z_4z_5=(z_3-1)(z_5-z_4)+z_4(z_3-z_5)~,~~~\eea
so that we can write $P_1-P_2-P_3=P_4+P_5$, with
\bea &&P_4=
-z_5(z_3-1)(z_4-1)(z_6-1)(z_3-z_6)(z_4-z_5)\prod_{i=3}^{6}z_i(
z_i-1) \prod_{3\leq i<j\leq 6}^{i\neq 5}(z_i - z_j)~,~~~\\
&&P_5=z_4z_5(z_4-1)(z_6-1)(z_3-z_5)(z_3-z_6)\prod_{i=3}^{6}z_i(
z_i-1) \prod_{3\leq i<j\leq 6}^{i\neq 5}(z_i - z_j)~,~~~\eea
which correspond to two-cycles
\bea  \{(1,2,7,4,6,5,3),(1,2,5,6,7,3,4)\}~~,~~
\{(1,2,3,4,5,6,7),(1,3,7,2,5,6,4)\}~~~~\label{twocycle3}\eea
respectively with $\Tr(P'_4Q'^{-1})=0$, $\Tr(P'_5Q'^{-1})=0$.
%
%

We thus conclude that strictly speaking, the amplitude-level identity between
(\ref{twocycle1}) and (\ref{twocycle2}) is up to some CHY-integrands
which have vanishing amplitude.
More explicitly, the identity
(\ref{twocycle1})=(\ref{twocycle2})+(\ref{twocycle3}) holds exactly
at the integrand-level, while (\ref{twocycle3}) has vanishing final
result, so that (\ref{twocycle1})=(\ref{twocycle2}) holds at the amplitude-level.
 This provides the amplitude-level identity an
explanation from the basic linearity of the trace.

\subsubsection{Yang-Mills theory}
For 5-point amplitude in Yang-Mills theory, we want to compute
\bea
\cA^{\YM}_5=\sum_{sol}{z_{12}^2z_{25}^2z_{51}^2\over
|\Phi|^{125}_{125}}{\pf' \Psi_{10\times 10}\over
  z_{12}z_{23}z_{34}z_{45}z_{51} }\equiv
\sum_{z_3,z_4 \in sol} {P^{\YM}(z_3,z_4)\over Q^{\YM}(z_3,z_4)}~,~~~
\eea
under gauge-fixing $z_1=\infty, z_2=1,z_5=0$. Let us choose the
reduced Pfaffian as
\bea \pf' \Psi_{10\times 10 }={-2\over z_1-z_2}\pf
\Psi^{12}_{12}~,~~~\eea
where $\Psi^{12}_{12}$ is a $8\times 8$ matrix. As in the 4-point
case, the large $z_1$ dependence of $\pf' \Psi_{10\times 10}$ is
$1/z_1^2$, while $1/(z_{12}z_{23}z_{34}z_{45}z_{51})$ is also
$1/z_1^2$. Together with the factor $z_{12}^2z_{25}^2z_{51}^2$ in
numerator, we get a finite integrand under the $z_1\to \infty$ limit.

We now follow the standard computation procedure:
\begin{enumerate}
\item Write down the
expressions for $|\Phi|^{125}_{125}$ and $\pf'\Psi_{10\times 10}$,
and work out $P^{\YM}(z_3,z_4)$, $Q^{\YM}(z_3,z_4)$;
\item Replace the
variables $z_i$'s by companion matrices $T_{z_i}$, as $P'=P|_{z_i\to
  T_{z_i}}$, $Q'=Q|_{z_i\to T_{z_i}}$;
\item Compute the inverse of $Q'$ and the trace $\Tr(P'Q'^{-1})$.
\end{enumerate}

The result for un-specified helicities is quite lengthy. For
illustration, let us consider the 5-point amplitude with helicity
$\cA^{\YM}_5(g_1^-,g_2^-,g_3^+,g_4^+,g_5^+)$. The polarization
vector is defined as
\bea \epsilon_{\mu}^-(k)={\spab{k|\gamma_\mu|r}\over
\sqrt{2}\spbb{k~r}}~~~,~~~\epsilon_{\mu}^{+}(k)={\spab{r|\gamma_{\mu}|k}\over
\sqrt{2}\spaa{r~k}}~~,~~~\eea
and we choose the reference momenta as $r_1=r_2=k_3$,
$r_3=r_4=r_5=k_2$. Thus settled, the only surviving products of
polarization vectors are $\epsilon^-(k_1)\cdot \epsilon^+(k_4)$ and
$\epsilon^-(k_1)\cdot \epsilon^+(k_5)$. After imposing momentum
conservation for $\epsilon^{\pm}(k_i)\cdot k_j$ to reduce the
ambiguity, we can simplify the $8\times 8$ matrix
$\widetilde{\Psi}\equiv\Psi^{12}_{12}$ as
\bea \left(
\begin{array}{cccccccc}
 0 & \frac{k_3 k_4}{z_3-z_4} & \frac{k_3 k_5}{z_3-z_5} & 0 & 0 & \widetilde{\Psi}_{16} & -\frac{\epsilon_4 k_3}{z_4-z_3} & -\frac{\epsilon_5 k_3}{z_5-z_3} \\
 \frac{k_3 k_4}{z_4-z_3} & 0 & \frac{k_4 k_5}{z_4-z_5} & -\frac{\epsilon_1 k_4}{z_1-z_4} & -\frac{\epsilon_2 k_4}{z_2-z_4} & -\frac{\epsilon_3 k_4}{z_3-z_4} & \widetilde{\Psi}_{27} & \frac{\epsilon_5 k_1+\epsilon_5 k_3}{z_5-z_4} \\
 \frac{k_3 k_5}{z_5-z_3} & \frac{k_4 k_5}{z_5-z_4} & 0 & \frac{\epsilon_1 k_2+\epsilon_1 k_4}{z_1-z_5} & \frac{\epsilon_2 k_1+\epsilon_2 k_4}{z_2-z_5} & \frac{\epsilon_3 k_1+\epsilon_3 k_4}{z_3-z_5} & \frac{\epsilon_4 k_1+\epsilon_4 k_3}{z_4-z_5} & \widetilde{\Psi}_{38} \\
 0 & \frac{\epsilon_1 k_4}{z_1-z_4} & -\frac{\epsilon_1 k_2+\epsilon_1 k_4}{z_1-z_5} & 0 & 0 & 0 & \frac{\epsilon_1 \epsilon_4}{z_1-z_4} & \frac{\epsilon_1 \epsilon_5}{z_1-z_5} \\
 0 & \frac{\epsilon_2 k_4}{z_2-z_4} & -\frac{\epsilon_2 k_1+\epsilon_2 k_4}{z_2-z_5} & 0 & 0 & 0 & 0 & 0 \\
 \widetilde{\Psi}_{61}& \frac{\epsilon_3 k_4}{z_3-z_4} & -\frac{\epsilon_3 k_1+\epsilon_3 k_4}{z_3-z_5} & 0 & 0 & 0 & 0 & 0 \\
 \frac{\epsilon_4 k_3}{z_4-z_3} & \widetilde{\Psi}_{72} & -\frac{\epsilon_4 k_1+\epsilon_4 k_3}{z_4-z_5} & \frac{\epsilon_1 \epsilon_4}{z_4-z_1} & 0 & 0 & 0 & 0 \\
 \frac{\epsilon_5 k_3}{z_5-z_3} & -\frac{\epsilon_5 k_1+\epsilon_5 k_3}{z_5-z_4} & \widetilde{\Psi}_{83} & \frac{\epsilon_1 \epsilon_5}{z_5-z_1} & 0 & 0 & 0 & 0 \\
\end{array}
\right)~,~~~ \eea
where
\bea &&\widetilde{\Psi}_{16}=-\widetilde{\Psi}_{61}=\frac{\epsilon_3
k_1}{z_3-z_1}+\frac{\epsilon_3 k_4}{z_3-z_4}-\frac{\epsilon_3
k_1+\epsilon_3 k_4}{z_3-z_5}~,~~~ \nonumber\\
&&\widetilde{\Psi}_{27}=-\widetilde{\Psi}_{72}=\frac{\epsilon_4
k_1}{z_4-z_1}+\frac{\epsilon_4 k_3}{z_4-z_3}-\frac{\epsilon_4
k_1+\epsilon_4 k_3}{z_4-z_5}~,~~~\nonumber\\
&&\widetilde{\Psi}_{38}=-\widetilde{\Psi}_{83}=\frac{\epsilon_5
k_1}{z_5-z_1}+\frac{\epsilon_5 k_3}{z_5-z_3}-\frac{\epsilon_5
k_1+\epsilon_5 k_3}{z_5-z_4}~.~~~\eea
This greatly simplifies the result of reduced Pfaffian, which
reads, after our gauge-fixing,
\bea \pf'\Psi_{10\times 10}\equiv{N_{\Psi}\over D_{\Psi}} &=&-2{z_4
\kappa_{2, 1} +
    z_4 \kappa_{2, 4}-\kappa_{2, 1} \over z_3z_4(z_4-1)(z_3 - z_4) }
   \Big( \big(z_3-z_4\big) \widetilde{\epsilon}_{1, 5} \kappa_{3, 1} \kappa_{4, 1}- \big(z_3-z_4\big) \widetilde{\epsilon}_{1, 4} \kappa_{3, 1} \kappa_{5, 1}\nonumber\\
    && -
    (z_3-z_4) \widetilde{\epsilon}_{1, 4} \kappa_{3, 1} \kappa_{5, 3}
  - z_4 \widetilde{\epsilon}_{1, 5} \kappa_{3, 4} \kappa_{4, 1} +
    z_3 \widetilde{\epsilon}_{1, 5} \kappa_{3, 1} \kappa_{4, 3}  + z_4 \widetilde{\epsilon}_{1, 4} \kappa_{3, 4} \kappa_{5, 1}
    \Big)~,~~~\label{A5YMPf}
 \eea
where we recall again that
$\widetilde{\epsilon}_{i,j}=\epsilon_i\epsilon_j$,
$\kappa_{i,j}=\epsilon_ik_j$. The factor of scalar part
$1/(z_{12}z_{23}z_{34}z_{45}z_{51})$ after gauge fixing is ${1\over
z_4(z_3-1) (z_3 - z_4) }$, and the Jacobian $|\Phi|^{125}_{125}$ is
the same as in the scalar theory,
\bea {1\over |\Phi|^{125}_{125}}={z_3^2z_4^2(z_3-1)^2( z_4-1)^2 (z_3
- z_4)^2 \over Q^{\YM} }~,~~~\eea
where $Q^{\YM}$ is a polynomial of $z_3,z_4$ and Mandelstam
variables, and it is also the denominator of integrand. The
numerator of $1/|\Phi|^{125}_{125}$ cancels the denominator of
$\pf'\Psi$ and that of scalar part, leaving a factor $z_3( z_3-1) (
z_4-1)$ in the numerator. Combined with the numerator $N_{\Psi}$ of
$\pf'\Psi_{10\times 10}$, they contribute to $P^{\YM}=z_3(
z_3-1) ( z_4-1)N_{\Psi}$.

Then it is straightforward to apply the replacements
$P'^{\YM}(T_{z_3},T_{z_4})=P^{\YM}(z_3,z_4)|_{z_i\to T_{z_i}}$,
$Q'^{\YM}(T_{z_3},T_{z_4})=Q^{\YM}(z_3,z_4)|_{z_i\to T_{z_i}}$, and
compute the trace $\Tr(P'^{YM}(Q'^{\YM})^{-1})$. To make the
computation more efficient, we can firstly apply the polynomial
reduction of $P^{\YM}(z_3,z_4)$, $Q^{\YM}(z_3,z_4)$ over $\gb(I)$.
The remainders $R(P^{\YM}), R(Q^{\YM})$ are polynomials of $z_{4}$
only, since the monomial basis is $\{1,z_4\}$. Then we can proceed
by replacing $z_4\to T_{z_4}$ for the remainders, and compute the
corresponding trace. This gives the same result as with the original
$P^{\YM}, Q^{\YM}$, but the computation would be much faster. With
Mathematica, we obtain
\bea &&\Tr(P'^{\YM}(Q'^{\YM})^{-1})=
 2{ \widetilde{\epsilon}_{1, 5} \kappa_{2, 1} \kappa_{3, 1} \kappa_{4, 3} +
  \widetilde{\epsilon}_{1, 4} \kappa_{2, 1} \kappa_{3, 4} \kappa_{5, 1}
 - \widetilde{\epsilon}_{1, 5} \kappa_{2, 1} \kappa_{3, 4} \kappa_{4,
 1}\over s_{12}s_{34}}\nonumber\\
&& ~~~~~~+2{ \widetilde{\epsilon}_{1, 5} \kappa_{2, 1} \kappa_{3, 1}
\kappa_{4, 1} +
  \widetilde{\epsilon}_{1, 5} \kappa_{2, 1} \kappa_{3, 1} \kappa_{4, 3} -
  \widetilde{\epsilon}_{1, 4} \kappa_{2, 1} \kappa_{3, 1} \kappa_{5, 1} -
  \widetilde{\epsilon}_{1, 4} \kappa_{2, 1} \kappa_{3, 1} \kappa_{5,
  3}\over s_{12}s_{45}}\nonumber\\
&&~~~~~~~~~~~~~~~~~~~~~~~~~~~~+2{ \widetilde{\epsilon}_{1, 5}
\kappa_{2, 4} \kappa_{3, 4} \kappa_{4, 1} -
  \widetilde{\epsilon}_{1, 5} \kappa_{2, 4} \kappa_{3, 1} \kappa_{4, 3} -
  \widetilde{\epsilon}_{1, 4} \kappa_{2, 4} \kappa_{3, 4} \kappa_{5,
  1}\over s_{15}s_{34}}~.~~~\eea

 The missing of the pole terms ${1/ (s_{15}s_{23})}, {1/(s_{23}s_{45})}$
 (terms involving pole $s_{23}$) is due to the choice of polarization vectors.
 However, the $s_{23}$ pole do exist, hiding
 in $\kappa_{2,i}, \kappa_{3,i}$.
 Directly rewriting the spinor brackets for $\widetilde{\epsilon}_{i,j},\kappa_{i,j}$ and $s_{ij}$, and using the Schouten identities 
 we get the famous MHV-amplitude \cite{Parke:1986gb, Berends:1987me} 
\bea \Tr(P'^{\YM}(Q'^{\YM})^{-1})={\spaa{1~2}^4\over
\spaa{1~2}\spaa{2~3}\spaa{3~4}\spaa{4~5}\spaa{5~1}}~.~~~\eea
%

\subsubsection{Gravity and $n$-point KLT Relations}
For 5-point amplitude in pure gravity theory, under gauge-fixing
$z_1=\infty$, $z_2=1$, $z_5=0$, we wish to compute
\bea
\cA^{\g}_5=\sum_{sol}{z_{12}^2z_{25}^2z_{51}^2\over
|\Phi|^{125}_{125}}(\pf'\Psi_{10\times
10})(\pf'\widetilde{\Psi}_{10\times
10})=\sum_{sol}{z_{12}^2z_{25}^2z_{51}^2\over
  |\Phi|^{125}_{125}}(\pf'\Psi_{10\times 10})^2~.~~~
\eea
Let us consider the gravity amplitude
$\cA_5^{\g}(1^{--},2^{--},3^{++},4^{++},5^{++})$, so that we can use
the same reduced Pfaffian $\pf'\Psi_{10\times 10}$ as in the
Yang-Mills case. Here, we do not have the factor of scalar part, but
the square of the factor of the reduced Pfaffian. The numerator of
$1/|\Phi|^{125}_{125}$ cancels the squared denominator of reduced
Pfaffian $z_3^2z_4^2(z_4-1)^2(z_3-z_4)^2$, leaving a factor of
$(z_3-1)^2$ in the numerator. Hence, we have $Q^{\g}=Q^{\YM}$, and
$P^{\g}=(z_3-1)^2N_{\Psi}^2$ with $N_{\Psi}$ given in
(\ref{A5YMPf}).

Thus, all the ingredients have been computed
in the Yang-Mills situation above, and we only need to work out the trace
$\Tr(P'^{\g}(Q'^{\g})^{-1})$, which gives a lengthy result:
{\small
\bea {\Spaa{1~2}^4\Big(\Spaa{1~2}^7 \Spaa{1~5} \Spaa{3~4}
\Spbb{2~1}^4 \Spbb{3~1}^3 \Spbb{4~2} \Spbb{4~3}^2
\Spbb{5~1}^3+~\mbox{971~more~terms}\Big)\over \Spaa{1~3} \Spaa{1~4}
\Spaa{1~5} \Spaa{2~3}^2 \Spaa{2~4}^2 \Spaa{2~5}^3 \Spaa{3~4}
\Spaa{3~5} \Spaa{4~5} \Spbb{2~1} \Spbb{3~1}^2 \Spbb{3~2}^2
\Spbb{4~1} \Spbb{4~2} \Spbb{5~1} \Spbb{5~2} \Spbb{5~3}
\Spbb{5~4}}\nonumber\eea}
where we can see that all poles $s_{i,j},i,j=1,\ldots, 5$ appearing
therein, indicating the colorless structure of gravity amplitude.

This complicated expression can be simplified by non-trivially
imposing momentum conservation and Schouten identities. Applying the
algorithm described in the appendix of \cite{Feng:2015qna} ,
for instance, we can simplify
$A^{\g}_5(1^{--},2^{--},3^{++},4^{++},5^{++})$ to
\bea&&\frac{\Spaa{1~2}^6 \Spbb{4~3} \Spbb{5~3}}{\Spaa{1~4}
\Spaa{1~5} \Spaa{2~4} \Spaa{2~5} \Spaa{3~4}
\Spaa{3~5}}+\frac{\Spaa{1~2}^6 \Spbb{4~3} \Spbb{5~4}}{\Spaa{1~3}
\Spaa{1~5} \Spaa{2~3} \Spaa{2~5}
\Spaa{3~4} \Spaa{4~5}}\nonumber\\
&&+\frac{\Spaa{1~2}^6 \Spbb{5~3} \Spbb{5~4}}{\Spaa{1~3} \Spaa{1~4}
\Spaa{2~3} \Spaa{2~4} \Spaa{3~5} \Spaa{4~5}}~.~~~\eea
which agrees perfectly with the result given by KLT relation \cite{Kawai:1985xq, Bern:1998sv, BjerrumBohr:2010ta, BjerrumBohr:2010zb, BjerrumBohr:2010yc}.

%
%

More generally, for $n$-point amplitude, under the usual gauge-fixing
$z_1=\infty$, $z_2=1$, $z_n=0$, we wish to compute
\bea
\cA^{\g}_n=\sum_{sol}{z_{12}^2z_{2n}^2z_{n1}^2\over
|\Phi|^{12n}_{12n}}(\pf'\Psi_{2n\times
2n})(\pf'\widetilde{\Psi}_{2n\times
2n})=\sum_{sol}{z_{12}^2z_{2n}^2z_{n1}^2\over
|\Phi|^{12n}_{12n}}(\pf'\Psi_{2n\times 2n})^2~.~~~\eea
In order to write down the reduced Pfaffian, we need to compute the
Pfaffian of a $(2n-2)\times (2n-2)$ matrix, which is quite complicated.
Direct computation using the above formula is obviously very difficult,
just like the direct computation of gravity amplitude by Feynman
diagram. So we would like to follow the KLT formalism, and compute
the gravity amplitude as square of Yang-Mills amplitudes.

An important property of the reduced Pfaffian is that, it can be
expanded \cite{Cachazo:2013iea} as
\bea \pf'\Psi=\sum_{\alpha\in S_{n-3}}{\sum_{\beta\in
S_{n-3}}S[\alpha|\beta]A_{n}^{\YM}(1,\beta,n,n-1)\over
(z_{1}-z_{\alpha_2})(z_{\alpha_2}-z_{\alpha_3})\cdots
(z_{\alpha_{n-2}}-z_{n-1})(z_{n-1}-z_{n})(z_n-z_1)}~,~~~\eea
where $\alpha,\beta$ are permutations of labels $2,3,\ldots, n-2$,
and $S[\alpha|\beta]$ is the S-kernel. The appearance of
$\cA_n^{\YM}$ is a consequence of certain integrand summing over all
$(n-3)!$ solutions of scattering equations in the original
derivation, and in the companion matrix method, it corresponds to
the trace of that integrand when changing variables to companion
matrices. In any event, it is a constant, and can be dragged out of
the trace.

Using this expression, we can expand one $\pf'\Psi$ in the gravity amplitude,
\bea && \cA^{\g}_n=\sum_{sol}\sum_{\alpha\in
S_{n-3}}\left({P(z_3,z_4,\ldots,z_{n-1})\over Q(z_3,z_4,\ldots,
z_{n-1})}\right)\times\sum_{\beta\in
  S_{n-3}}S[\alpha|\beta]
\cA_{n}^{\YM}(1,\beta,n,n-1)~,~~~\\
&&{P\over Q}\equiv {z_{12}^2z_{2n}^2z_{n1}^2\over
|\Phi|^{12n}_{12n}}{\pf'\Psi_{2n\times 2n}\over
z_{1\alpha_2}z_{\alpha_2\alpha_3}\cdots
z_{\alpha_{n-2},n-1}z_{n-1,n}z_{n1} }~.~~~ \eea
The trace $\Tr(P'(T_{z_i})Q'^{-1}(T_{z_i}))$ for the set $\alpha$
gives $\cA_n^{\YM}(1,\alpha,n-1,n)$, and the summation over
permutations of $\alpha$ can be taken out of the trace, and we
thereby arrive at the KLT relation\footnote{Note that the ordering
of set $\alpha$(or $\beta$) here defined in \cite{Cachazo:2013iea}
is the reverse of that defined in \cite{BjerrumBohr:2010ta}. }.


\subsection{Six-Point Amplitudes}
We proceed onto six-point amplitudes, i.e., $n=6$.
Using the standard gauge-fixing $z_1=\infty$, $z_2=1$, $z_6=0$,
Dolan-Goddard's polynomial form \cite{Dolan:2014ega} of the scattering equations is given by
\bea
&&f_1=s_{12}+s_{13}z_{3}+s_{14}z_4+s_{15}z_5~,~~~\\
&&f_2=s_{123}z_3+s_{124}z_4+s_{125}z_5
+s_{134}z_3z_4+s_{135}z_3z_5+s_{145}z_4z_5~,~~~\\
&&f_3=s_{56}z_3z_4+s_{46}z_3z_5+s_{36}z_4z_5+s_{26}z_3z_4z_5~.~~~\eea
We can thus define the ideal $I=\ideal{f_1,f_2,f_3}$ in the polynomial ring
$\mathbb{C}[z_3,z_4,z_5]$.
The degree of ideal $I$ is 6, so according to B\'ezout's theorem, it has 6 solutions, though it is not possible to obtain analytic expressions for these solutions, as already seen in the 5-point
cases.
Let us then consider the companion matrix method.

We generate the Gr\"obner basis for $I$ in {\sl Lexicographic}
ordering $z_3\prec z_4\prec z_5$. Analytically, the explicit
expression of $\gb(I)$ is rather complicated, especially in the
presence of so many parameters $s_{ij}$ in the ring. By varying the
exponents to some high power, the polynomial reduction of the
monomials $z_3^{a_3}z_4^{a_4}z_5^{a_5}$ (with $a_i$ from 0 to some
finite number, say 20) over $GB(I)$ gives the monomial basis
\bea \B=\{1,z_5,z_5^2,z_5^3,z_5^4,z_5^5\}~.~~~\nonumber\eea
The polynomial reduction of $z_3\B, z_4\B$ and $z_5\B$ over $\gb(I)$
gives the companion matrices $T_{z_3},T_{z_4},T_{z_5}$, which are
$6\times 6$ matrices. Again, we need to compute $P'=P|_{z_3\to
T_{z_3},z_4\to T_{z_4},z_5\to T_{z_5}}$ $Q'=Q|_{z_3\to
T_{z_3},z_4\to T_{z_4},z_5\to T_{z_5}}$, and the final amplitude is
given by $\cA_6=\Tr(P'Q'^{-1})$, without summing over all solutions of
scattering equations.

Since the operations we need are
multiplication of matrices, taking inverse or trace of matrices, so
in principle it can be done analytically. However, the symbolic
manipulation for $n=6$ case is quite complicated, especially when
taking the inverse of matrix $Q'$ and simplifying the tedious trace
result in Mathematica, so we introduce random numeric kinematics -- i.e., by Monte Carlo assignments of the parametres $s_{ij}$ -- to get the final result.
One will see that, as is customary with coefficient fields in polynomial rings, trying a few large prime numbers would suffice very quickly.

\subsubsection{Scalar $\phi^3$ theory}
\def\I{\i}
We can write the amplitude as
\bea
\cA_6&=&\sum_{sol}{z_{12}^2z_{26}^2z_{61}^2\over
|\Phi|^{126}_{126}}{1\over
z_{12}^2z_{23}^2z_{34}^2z_{45}^2z_{56}^2z_{61}^2}\nonumber\\
&=&\sum_{sol}{1\over
|\Phi|^{126}_{126}(z_3-1)^2(z_3-z_4)^2(z_4-z_5)^2z_5^2}\equiv
\sum_{z_3,z_4,z_5 \in sol} {P(z_3,z_4,z_5)\over Q(z_3,z_4,z_5)}~,~~~
\eea
where
\bea \Phi^{126}_{126}=\left(
\begin{array}{ccc}
  \Phi_{33} & \Phi_{34} & \Phi_{35} \\
  \Phi_{43} & \Phi_{44} & \Phi_{45} \\
  \Phi_{53} & \Phi_{54} & \Phi_{55} \\
\end{array}
\right)~~~,~~~|\Phi|^{126}_{126}=\Det(\Phi^{126}_{126})~.~~~\eea

\paragraph{Prime Kinematic Strategy: }
The idea is the following.
Since we know for scalar $\phi^3$ theory, the final result of $\cA_6$ should take the form
\bea
\cA_6=\sum_{\I,\I_i}{c_{\I}\over s_{\I_1}s_{\I_2}s_{\I_3}}~,~~~
\eea
where $s_{\I_i}$ are the independent Mandelstam variables of
physical poles $s_{12}$, $s_{23}$, $s_{34}$, $s_{45}$, $s_{56}$,
$s_{16}$, $s_{123}$, $s_{234}$, $s_{345}$, and the summation is over
all possible products of three physical poles, e.g., ${1\over
  s_{12}s_{23}s_{56}}$, ${1\over s_{12}s_{45}s_{234}}$, etc.
So in total we have $\binom{9}{3}=84$ terms, which we denote as $\mathcal{S}_{\I}$, $\I=1,2,\ldots 84$, and the amplitude is expanded as
$\cA_6=\sum_{\I=1}^{84}c_{\I}\mathcal{S}_{\I}$, where $c_{\I}$ is
either 0 or 1.

To each physical pole we now randomly assign a prime number, i.e., we are working with the much simpler polynomial ring $\IC[z]$ instead of $\IC(s)[z]$.
In this case, the computation of $\Tr(P'Q'^{-1})$ is trivial
within seconds, and the result as well as $\mathcal{S}_{\I}$'s are
all numbers.
Next, we shall find the solutions
$\sum_{\I=1}^{84}c_{\I}\mathcal{S}_{\I}=\Tr(P'Q'^{-1})$ for $c_{\I}$'s.
However, doing this by brute-force is impossible since there are 84 $c_{\I}$'s and each one can take $0$ or $1$, so one would go through all $2^{84}$
possibilities, which is far beyond any computational ability.

We therefore adopt the following strategy: instead of setting all coefficients to numbers, we can assign all physical poles to prime numbers
except one pole.  For example, we would leave $s_{345}$, to detect first the
coefficients of $\mathcal{S}_{\I}$'s which contains the pole $1\over
s_{345}$. Keeping one symbolic variable $s_{345}$ would extend the
computation time of $\Tr(P'Q'^{-1})$ up to minutes, but it is still very
manageable, while keeping two or more symbolic variables would make
the computation of $\Tr(P'Q'^{-1})$ in Mathematica very hard for a
laptop.

Let us see the above strategy in action. Setting the kinematics
(coefficient variables) as, e.g.,
\bea
&&s_{12}=7~~,~~s_{23}=37~~,~~s_{34}=79~~,~~s_{45}=97~,~~~\nonumber\\
&&s_{56}=131~~,~~s_{16}=179~~,~~s_{123}=181~~,~~s_{234}=223~,~~~\nonumber\eea
while leaving $s_{345}$ free, we get
\bea \Tr(P'Q'^{-1})=-{64909247478\over 1878479042622679} -
{32736\over 9601739 s_{345}}~.~~~\eea
Among the $\mathcal{S}_{\I}$'s, there are $\binom{8}{2}=28$ terms
containing physical pole $s_{345}$, and the number marked by
${1\over s_{345}}$ in $\Tr(P'Q'^{-1})$ should be expanded into these
28 terms\footnote{In fact, using the compatibility among poles, we
can greatly reduce the number of possible combinations of poles. We
will consider this fact in latter examples.}. This is thus a problem
in {\it Egyptian fractions}. By going through all $2^{28}$
possibilities of $c_{\I}$, we find the unique expansion
\bea {32736\over 9601739 }{1\over s_{345}}=\Big({1\over 7\times 79
}+{1\over 79\times 179}+{1\over 7\times 97}+{1\over97\times
179}\Big){1\over s_{345}}~,~~~\eea
So mapping to the physical poles, we find that
\bea -\frac{1}{s_{12} s_{34} s_{345}}-\frac{1}{s_{16} s_{34}
s_{345}}-\frac{1}{s_{12} s_{45} s_{345}}-\frac{1}{s_{16} s_{45}
s_{345}}~~~~\label{phi3A6s345}\eea
is a part of $\cA_6$.

Now, we try to get more poles.
Taking the kinematics as, e.g.,
\bea
&&s_{12}=7~~,~~s_{23}=37~~,~~s_{34}=79~~,~~s_{45}=97~,~~~\nonumber\\
&&s_{56}=131~~,~~s_{16}=179~~,~~s_{123}=181~~,~~s_{345}=251~,~~~\nonumber\eea
while leaving $s_{234}$ free, we get
\bea \Tr(P'Q'^{-1})=-\frac{35960}{68541427
  s_{234}}-\frac{13829207594}{302048840293589}~.~~~
\eea
The part marked by the physical pole $s_{234}$ can be uniquely expanded
as
\bea \frac{35960}{68541427}{1\over s_{234}}=\Big({1\over 37\times
179}+{1\over 79\times 179}+{1\over 37\times 131}+{1\over 79\times
131}\Big){1\over s_{234}}~,~~~\eea
thus
\bea -\frac{1}{s_{16} s_{23} s_{234}}-\frac{1}{s_{16} s_{34}
s_{234}}-\frac{1}{s_{23} s_{56} s_{234}}-\frac{1}{s_{34} s_{56}
s_{234}}~,~~~\label{phi3A6s234}\eea
is also a part of $\cA_6$.
With the same procedure, we find that for
physical pole $s_{123}$,
\bea -\frac{1}{s_{12} s_{45} s_{123}}-\frac{1}{s_{23} s_{45}
s_{123}}-\frac{1}{s_{12} s_{56} s_{123}}-\frac{1}{s_{23} s_{56}
s_{123}}~~~~\label{phy3A6s123}\eea
is also part of $\cA_{6}$. Finally, we need to determine the coefficients
$c_{\I}$ of $\mathcal{S}_{\I}$'s without physical poles
$s_{123},s_{234},s_{345}$. There are in total $\binom{6}{3}=20$
terms. Taking the kinematics as, e.g.,
\bea
&&s_{12}=7~~,~~s_{23}=37~~,~~s_{34}=79~~,~~s_{45}=97~,~~~\nonumber\\
&&s_{56}=131~~,~~s_{16}=179~~,~~s_{123}=181~~,~~s_{234}=223~~,~~s_{345}=251~,~~~\nonumber
\eea
computing the $\Tr(P'Q'^{-1})$ and extracting the contributions from
results (\ref{phi3A6s345}), (\ref{phi3A6s234}), (\ref{phy3A6s123}),
the remaining result can be uniquely expanded as
\bea -\frac{714874}{46539628933}=-{1\over 7\times 79\times
131}-{1\over 37\times 97\times 179}~,~~~\eea
so the last part for $\cA_6$ is
\bea-{1\over s_{12}s_{34}s_{56}}-{1\over
s_{16}s_{23}s_{45}}~.~~~\eea

Putting all the above together, we therefore conclude that
\bea
\cA_6&=&-\Big(\frac{1}{s_{12} s_{34} s_{56}}+\frac{1}{s_{16}
s_{23} s_{45}}+\frac{1}{s_{12} s_{45} s_{123}}+\frac{1}{s_{23}
s_{45}
s_{123}}+\frac{1}{s_{12} s_{56} s_{123}}\nonumber\\
&&+\frac{1}{s_{23} s_{56} s_{123}}+\frac{1}{s_{12} s_{34}
s_{345}}+\frac{1}{s_{16} s_{34} s_{345}}+\frac{1}{s_{12} s_{45}
s_{345}}+\frac{1}{s_{16} s_{45} s_{345}}\nonumber\\
&&+\frac{1}{s_{16} s_{23} s_{234}}+\frac{1}{s_{16} s_{34}
s_{234}}+\frac{1}{s_{23} s_{56} s_{234}}+\frac{1}{s_{34} s_{56}
s_{234}}\Big)~.~~~\eea
This prime-numeric method can be applied to all the cases of $n=6$
amplitudes of scalar $\phi^3$ theory.

\subsubsection{Yang-Mills theory}

For Yang-Mills theory, when $n=6$, we meet the first "not so simple"
gluon amplitude, i.e., the next-MHV amplitude, so it is worthwhile
to verify the companion matrix method with this non-trivial example.
To illustrate, let us consider the split helicity amplitude
$\cA_6^{\YM}(g_1^-,g_2^-,g_3^-,g_4^+,g_5^+,g_6^+)$, and choose the
reference momenta as $r_1=r_2=r_3=k_4$, $r_4=r_5=r_6=k_3$, so that
only $\widetilde{\epsilon}_{1,5}$, $\widetilde{\epsilon}_{1,6}$,
$\widetilde{\epsilon}_{2,5}$, $\widetilde{\epsilon}_{2,6}$ are
non-zero. The object we want to compute is
\bea
\cA_6^{\YM}=\sum_{sol}{z_{12}^2z_{26}^2z_{61}^2\over
|\Phi|^{126}_{126}}{\pf'\Psi_{12\times 12}\over
  z_{12}z_{23}z_{34}z_{45}z_{56}z_{61}}\equiv
\sum_{z_3,z_4,z_5 \in sol}
    {P^{\YM}(z_3,z_4,z_5)\over Q^{\YM}(z_3,z_4,z_5)}~.~~~
    \eea
Here both the Jacobian $|\Phi|^{126}_{126}$ and reduced Pfaffian
$\pf'\Psi_{12\times 12}$ are very complicated, so it is almost
impossible to compute it analytically. As in the scalar $\phi^3$
example, we can follow the semi-analytic procedure, and set the
physical poles as some prime numbers, while keeping $\Psi_{12\times
12}$(all $\widetilde{\epsilon}_{i,j}=\epsilon_i\epsilon_j,
\kappa_{i,j}=\epsilon_ik_j$ and $k_ik_j$ in $\Psi$) analytic.
In
this case, the ideal and Gr\"obner basis are just algebraic systems
of polynomials with integer coefficients, while the elements of
companion matrices are rational numbers. So the computation is very
fast.

The Jacobian under the chosen gauge-fixing is
\bea {1\over
|\Phi|^{126}_{126}}={z_3^2z_4^2z_5^2(z_3-1)^2(z_4-1)^2(z_5-1)^2(z_3-z_4)^2(z_3-z_5)^2(z_4-z_5)^2\over
D_{\Phi}(z_3,z_4,z_5)}~,~~~\eea
where $D_{\Phi}$ is polynomial in $z_3,z_4,z_5$. The reduced
Pfaffian together with the factor of scalar part give
\bea
{N_{\Psi}(z_3,z_4,z_5,\widetilde{\epsilon}_{i,j},\kappa_{i,j},k_ik_j)\over
z_3z_4z_5^2(z_3-1)^2(z_4-1)( z_5-1)(z_3 - z_4)^2(z_3 - z_5)(z_4 -
z_5)^2 }
\eea
under the chosen gauge-fixing for some polynomial numerator $N_{\Phi}$.
So we have
\bea
P^{\YM}(z_3,z_4,z_5)={z_3z_4 ( z_4-1) ( z_5-1) (z_3 - z_5)
  N_{\Psi}}~~,~~
Q^{\YM}(z_3,z_4,z_5)=D_{\Phi}~.~~~\eea
Note that $N_{\Psi}$ originates from the Pfaffian of a $10\times 10$
antisymmetric matrix, where by definition, each term in the Pfaffian
is a product of five elements in the matrix. So each term in
$N_{\Psi}$ is a product of five elements selected from
$\widetilde{\epsilon}_{i,j}$, $\kappa_{i,j}$, $k_ik_j$, combined
with a monomial of $z_3,z_4,z_5$, for example, $2 z_3^3 z_4^4
\widetilde{\epsilon}_{2,6}\kappa_{1,2}\kappa_{3,1}\kappa_{4,1}\kappa_{5,1}$.

Finally we can take the replacement $P'^{\YM}=P^{\YM}|_{z_i\to
T_{z_i}}$, $Q'^{\YM}=Q^{\YM}|_{z_i\to T_{z_i}}$ and compute the
trace $\Tr(P'^{\YM}(Q'^{\YM})^{-1})$. It is given as
\bea \Tr(P'^{\YM}(Q'^{\YM})^{-1})&=&{44\over
6141149}\widetilde{\epsilon}_{2, 6}\kappa_{1, 2}\kappa_{3,
2}\kappa_{4, 1}\kappa_{5, 1}-{1\over 877307}\widetilde{\epsilon}_{2,
5}\kappa_{1, 3}\kappa_{3, 2}\kappa_{4, 1}\kappa_{6, 2}\nonumber\\
&&~~~~~~~~~~~~~~~~~~~~~~~~~~~~~~~~~~~~~~~~~~~~+500~\mbox{more~terms}~.~~~
\eea
Using the techniques shown in scalar $\phi^3$ theory, we can
uniquely decompose the rational numbers as
\bea {44\over 6141149}={1\over 7\times 181\times 131}+{1\over
181\times 37\times 131}~~,~~{1\over 877307}={1\over 181\times
37\times 131}~,~~~\eea
so we can conclude that
\bea \Tr(P'^{\YM}(Q'^{\YM})^{-1})&=&\Big( {1\over s_{12} s_{123}
s_{56}} +{ 1\over s_{123} s_{23} s_{56}}\Big)
\widetilde{\epsilon}_{2, 6}\kappa_{1, 2}\kappa_{3, 2}\kappa_{4,
1}\kappa_{5, 1}\nonumber\\
&&-{ 1\over s_{123} s_{23} s_{56}}\widetilde{\epsilon}_{2,
5}\kappa_{1, 3}\kappa_{3, 2}\kappa_{4, 1}\kappa_{6,
  2}+500~\mbox{more~terms}~.~~~
\eea
Rewriting them as spinor products and applying the simplification
algorithm for spinor expression, we get a one-page long result,
which remarkably agrees with the known answers
\cite{Kosower:2004yz,Luo:2005rx}.

\subsection{Seven-Point Amplitudes}
The companion matrices $T_{z_i}$ are simultaneously diagonalizable,
and according to Stickelberger's theorem, the complex roots $z_i$ of
ideal $I$ are the vectors of simultaneous eigenvalues of the
companion matrices $T_{z_i}$. Thus when they are evaluated in the
diagonal form, the matrices $P'=P|_{z_i\to T_{z_i}}$, $Q'=Q|_{z_i\to
T_{z_i}}$, $P'Q'^{-1}$ are also diagonal, and it builds the
one-to-one mapping between diagonal elements of $(n-3)!\times
(n-3)!$ matrix $P'Q'^{-1}$ and the integrand $P/Q$ evaluated at the
$(n-3)!$ complex solutions of scattering equations. To demonstrate
this, let us go through a 7-point example of scalar $\phi^3$ theory.

As usual, let us gauge fixing $z_1=\infty, z_2=1,z_7=0$, and the
amplitude is given by
\bea
\cA_7=\sum_{sol}{z_{12}^2z_{27}^2z_{71}^2\over
|\Phi|^{127}_{127}}{1\over
  z_{12}^2z_{23}^2z_{34}^2z_{45}^2z_{56}^2z_{67}^2z_{71}^2}
\equiv
\sum_{z_3,z_4,z_5,z_6 \in sol}{P(z_3,z_4,z_5,z_6)\over Q(z_3,z_4,z_5,z_6)}~,~~~
   \eea
where
\bea \Phi^{127}_{127}=\left(
\begin{array}{cccc}
  \Phi_{33} & \Phi_{34} & \Phi_{35} & \Phi_{36} \\
  \Phi_{43} &\Phi_{44} & \Phi_{45} & \Phi_{46} \\
  \Phi_{53} & \Phi_{54} & \Phi_{55} & \Phi_{56} \\
  \Phi_{63} & \Phi_{64} & \Phi_{65} & \Phi_{66} \\
\end{array}
\right)~~~,~~~|\Phi|^{127}_{127}=\Det(\Phi^{127}_{127})~.~~~
\eea
The Dolan-Goddard polynomial form \cite{Dolan:2014ega} of the scattering equations is given by
\bea &&f_1=s_{12}+s_{13}z_3+s_{14}z_{4}+s_{15}z_5+s_{16}z_6~,~~~\\
&&f_2=s_{123}z_3+s_{124}z_4+s_{125}z_5+s_{126}z_6\nonumber\\
&&~~~~~~+s_{134}z_3z_4+s_{135}z_3z_5+s_{136}z_3z_6+s_{145}z_4z_5+s_{146}z_4z_6+s_{156}z_5z_6~,~~~\\
&&f_3=s_{1234}z_3z_4+s_{1235}z_3z_5+s_{1236}z_3z_6+s_{1245}z_4z_5+s_{1246}z_4z_6+s_{1256}z_5z_6\nonumber\\
&&~~~~~~+s_{1345}z_3z_4z_5+s_{1346}z_3z_4z_6+s_{1356}z_3z_5z_6+s_{1456}z_4z_5z_6~,~~~\\
&&f_4=s_{67}z_3z_4z_5+s_{57}z_3z_4z_6+s_{47}z_3z_5z_6+s_{37}z_4z_5z_6+s_{27}z_3z_4z_5z_6~.~~~\eea
We can define the ideal $I=\ideal{f_1,f_2,f_3,f_4}$ in polynomial
ring $\mathbb{C}[z_3,z_4,z_5,z_6]$, and generate the Gr\"obner basis
of $I$ in {\sl Lexicographic} order $z_3\prec z_4\prec z_5\prec
z_6$. The degree of ideal $I$ is 24, so the variety of ideal $I$ is
given by 24 point solutions for which there are no closed form
solutions.

Let us set the kinematics (all physical poles) as some prime numbers,
\bea &&s_{12} = 5~,~ s_{23} = 37~,~ s_{34} = 43~,~ s_{45} = 61~,~
s_{56} =
97~,~ s_{67} = 101~,~ s_{17} = 139~,~~~ \nonumber\\
&&s_{123} = 151~,~ s_{234} = 163~,~ s_{345} = 191~,~ s_{456} =
211~,~ s_{567} = 223~,~ s_{671} = 251~,~ s_{712} = 263~~~~\label{primen7}
\eea
in the following computation. The solutions of scattering equations
$f_i=0,i=1,2,3,4$ requires computing the roots of equations of
degree 24, which has no closed form in radicals.
Doing it numerically, we get 24 solutions {\footnotesize
\bea &&sol_1:~z_3=20.9071~,~~z_4=1.66835~,~~z_5=7.08198~,~~z_6=-64.2332,\nonumber\\
&&sol_2:~z_3= 1.4223 - 0.318993 \mathbbm{i}~,~~z_4= 12.204 - 5.48743
\mathbbm{i}~,~~z_5= 0.342956 - 0.477119 \mathbbm{i}~,~~z_6=51.9097 -
32.886 \mathbbm{i},\nonumber\\
&&sol_3:~z_3= 1.4223 + 0.318993 \mathbbm{i}~,~~z_4= 12.204 + 5.48743
\mathbbm{i}~,~~z_5= 0.342956 + 0.477119 \mathbbm{i}~,~~z_6= 51.9097
+
  32.886 \mathbbm{i},\nonumber\\
&&sol_4:~z_3=27.2316~,~~z_4=1.76178~,~~z_5=13.0497~,~~z_6=-12.5157,\nonumber\\
&&sol_5:~z_3=1.34598~,~~z_4=-3.76733~,~~z_5=-1.28282~,~~z_6=-56.7763,\nonumber\\
&&sol_6:~z_3=4.92534 + 1.82303 \mathbbm{i}~,~~z_4= 2.04236 + 0.47052
\mathbbm{i}~,~~z_5= 0.12331 + 0.73366 \mathbbm{i}~,~~z_6=-36.88 -
1.74857
\mathbbm{i},~~~~~~~~~~~~~~~~~~~~~~~~~~~~~~~~~~~~~\nonumber\eea}
\vspace{-0.2in}
{\footnotesize \bea  &&sol_{7}:~z_3= 4.92534 - 1.82303
\mathbbm{i}~,~~z_4= 2.04236 - 0.47052 \mathbbm{i}~,~~z_5= 0.12331 -
0.73366 \mathbbm{i}~,~~z_6=-36.88 +
  1.74857 \mathbbm{i},\nonumber\\
&&sol_{8}:~z_3=-11.2804~,~~z_4=3.5116~,~~z_5=-6.80042~,~~z_6=-1.26394,\nonumber\\
&&sol_{9}:~z_3=1.19261~,~~z_4=-8.20104~,~~z_5=3.07784~,~~z_6=6.22689,\nonumber\\
&&sol_{10}:~z_3=1.18325 + 1.93745 \mathbbm{i}~,~~z_4= 0.29585 -
0.48639 \mathbbm{i}~,~~z_5= 0.56997 + 1.11008 \mathbbm{i}~,~~z_6=
0.15405 -
0.39359 \mathbbm{i},\nonumber\\
&&sol_{11}:~z_3=1.18325 - 1.93745 \mathbbm{i}~,~~z_4= 0.29585 +
0.48639 \mathbbm{i}~,~~z_5= 0.56997 - 1.11008 \mathbbm{i}~,~~z_6=
0.15405 +
  0.39359 \mathbbm{i},~~~~~~~~~~~~~~~~~~~~~~~~~~~~~~~~~~~~~\nonumber\\
&&sol_{12}:~z_3=-4.76521~,~~z_4=-3.05026~,~~z_5=-1.6908~,~~z_6=-0.488528,\nonumber\eea}
\vspace{-0.2in} {\footnotesize \bea
&&sol_{13}:~z_3= 0.576445~,~~z_4=-3.05135~,~~z_5=1.14498~,~~z_6=0.712806,\nonumber\\
&&sol_{14}:~z_3=1.78095 + 0.41639 \mathbbm{i}~,~~z_4= 2.1103 -
0.60663 \mathbbm{i}~,~~z_5= 0.52752 + 0.29927
\mathbbm{i}~,~~z_6=2.3283 -
1.39061 \mathbbm{i},~~~~\nonumber\\
&&sol_{15}:~z_3= 1.78095 - 0.41639 \mathbbm{i}~,~~z_4= 2.1103 +
0.60663 \mathbbm{i}~,~~z_5= 0.52752 - 0.29927 \mathbbm{i}~,~~z_6=
2.3283 +
  1.39061 \mathbbm{i},~~~~~~~~~~~~~~~~~~~~~~~~~~~~~~~~~~~~~~~~~~~~~~~~~~~\nonumber\\
&&sol_{16}:~z_3=1.86192~,~~z_4=0.877999~,~~z_5=0.795994~,~~z_6=0.601979,\nonumber\\
&&sol_{17}:~z_3=1.65547~,~~z_4=1.9848~,~~z_5=0.493798~,~~z_6=2.31186,\nonumber\\
&&sol_{18}:~z_3=0.327576~,~~z_4=-0.0855936~,~~z_5=0.212916~,~~z_6=-0.0559545,\nonumber\eea}
\vspace{-0.2in}
{\footnotesize\begin{eqnarray}
&&sol_{19}:~z_3=0.307828~,~~z_4= 0.645287~,~~z_5=0.0420044~,~~z_6=0.46483,~~~~~~~~~~~~~~~~~~~~~~~~~~~~~~~~~~~~~~~~~~~~~~~~~~~~~~~~~~~~~~~~~~~~~~~~~~~~~~~~~~~~~~~~~~~~~~~~~~~~~~~~~~~~~\nonumber\\
&&sol_{20}:~z_3=0.174313~,~~z_4=0.120642~,~~z_5=0.0855984~,~~z_6=0.0445606,\nonumber\\
&&sol_{21}:~z_3=0.031819~,~~z_4=0.15022~,~~z_5=0.00455376~,~~z_6=0.0545382,\nonumber\\
&&sol_{22}:~z_3=0.0191033~,~~z_4=0.0765079~,~~z_5=0.0145344~,~~z_6=0.00921803,\nonumber\\
&&sol_{23}:~z_3=-0.100486~,~~z_4=-0.0950558~,~~z_5=-0.00857861~,~~z_6=-0.0892275,\nonumber\\
&&sol_{24}:~z_3=-0.0162083~,~~z_4=0.0167369~,~~z_5=0.00970032~,~~z_6=-0.0265855~,\nonumber
\end{eqnarray} }
and the integrand summing over all solutions is given by
\bea \sum_{sol_i, \ i=1}^{24}{P(z_3,z_4,z_5,z_6)\over
Q(z_3,z_4,z_5,z_6)}=1.99605\times 10^{-6}~.~~~\eea

Let us now turn to the companion matrix method. The monomial basis
over $\gb(I)$ is given by 24 elements
\bea
\B&=&\{1,~z_6,~z_6^2,~z_6^3,~z_6^4,~z_6^5,~z_6^6,~z_6^7,~z_6^8,~z_6^9,~z_6^{10},~z_6^{11},~z_6^{12}~,~~~\nonumber\\
&&~~~~z_6^{13},~z_6^{14},~z_6^{15},~z_6^{16},~z_6^{17},~z_6^{18},~z_6^{19},~z_6^{20},~z_6^{21},~z_6^{22},~z_6^{23}
\}~.~~~\eea
Accordingly, by polynomial reduction of $z_i\B,i=3,4,5,6$ over
$\gb(I)$, we can get the companion matrices $T_{z_i},i=3,4,5,6$,
which are $24\times 24$ matrix and satisfying $T_{z_i}\B=z_i\B$. In
order to compute $\cA_7$, we can proceed as usual by computing
$\Tr(P|_{z_i\to T_{z_i}}Q|_{z_i\to T_{z_i}}^{-1})$, and the result
is
\bea
\Tr(P'Q'^{-1})=\frac{19260317055974762778118}{9649229470008137021319652355}\approx
1.99605\times 10^{-6}~,~~~\eea
which agrees with the numeric result given by summing over all
solutions of scattering equations. Again we see that, since the
computation only involves basic manipulations on matrix, we are able
to get the closed form result, and show that the final result is
rational functions of Mandelstam variables.

The companion matrices are simultaneously diagonalizable. We can
choose $T_{z_6}$ and compute its eigenvectors, since $T_{z_6}$ is
the simplest companion matrix by definition. Such computation
involves finding the roots of equations of degree 24, which
prohibits analytic solution. Now, $T_{z_6}$ has 24 column
eigenvectors $u_i,i=1,\ldots, 24$, and from them we can define the
transformation matrix $U=(u_1,\ldots, u_{24})_{24\times 24}$. Then
$T^{d}_{z_i}=U^{-1}T_{z_i}U,i=3,4,5,6$ are all diagonal matrices,
explicitly given as
\def\diag{\mbox{diag}}
{\footnotesize \bea \diag (T^d_{z_3})& =&\{20.9071~,~~ 1.42230 +
0.31899 \mathbbm{i}~,~~ 1.42230 - 0.31899 \mathbbm{i}~,~~
1.345982~,~~
 4.92534 - 1.82303 \mathbbm{i}~,~~~~~~~~~~~~~~~~~~~~~~~~~~~~~~~~~~~~~~~~~~~~~\nonumber\\
 && 4.92534 + 1.82303 \mathbbm{i}~,~~ 27.2316~,~~ 1.192609~,~~
 1.78095 - 0.41639 \mathbbm{i}~,~~
 1.78095 +
  0.41639 \mathbbm{i}~,~~ \nonumber\\
  &&1.65547~,~~ -11.28035~,~~ 0.576445~,~~ 1.86192~,~~ -4.76521~,~~
0.307828~,~~\nonumber\\
&& 1.18325 - 1.93745 \mathbbm{i}~,~~
 1.18325 +
  1.93745 \mathbbm{i}~,~~ -0.1004864~,~~ 0.327576~,~~ 0.0318190~,~~\nonumber\\
  && 0.174313~,~~ -0.0162082~,~~
0.0191033\}~,~~~\nonumber\eea}
\vspace{-0.15in}
{\footnotesize \bea \diag (T^d_{z_4})& =&\{1.66835~,~~ 12.20402 +
5.48743 \mathbbm{i}~,~~ 12.20402 - 5.48743 \mathbbm{i}~,~~
-3.76733~,~~
 2.04236 - 0.47052 \mathbbm{i}~,~~~~~~~~~~~~~~~~~~~~~~~~~~~~~~~~~~~~~~~~~~~~~~~~~~~~~~~~~~~~~~\nonumber\\
 && 2.04236 + 0.47052 \mathbbm{i}~,~~ 1.76178~,~~ -8.20104~,~~
 2.11030 + 0.60663 \mathbbm{i}~,~~
 2.11030 -
  0.60663 \mathbbm{i}~,~~\nonumber\\
 &&1.98480~,~~ 3.51160~,~~ -3.05135~,~~ 0.877999~,~~ -3.05026~,~~
 0.645287~,~~\nonumber\\
 &&
  0.295854 + 0.486386 \mathbbm{i}~,~~
 0.295854 -
  0.486386 \mathbbm{i}~,~~ -0.0950558~,~~ -0.0855936~,~~ 0.150220~,~~\nonumber\\
  && 0.1206420~,~~ 0.0167369~,~~
0.0765079\}~,~~~\nonumber\eea}
\vspace{-0.15in}
{\footnotesize \bea \diag (T^d_{z_5})& =&\{7.08198~,~~ 0.342956 +
0.477119 \mathbbm{i}~,~~ 0.342956 - 0.477119 \mathbbm{i}~,~~
-1.282815~,~~
 0.123310 - 0.733658 \mathbbm{i}~,~~~~~~~~~~~~~~~~~~~~~~~~~~~~~~~~~~~~~~~~~~~~~~~~~~~~~~~~~~~~~~~~~~~~~~~~~~\nonumber\\
 &&0.123310 + 0.733658 \mathbbm{i}~,~~ 13.04970~,~~ 3.07784~,~~
 0.527517 - 0.299273 \mathbbm{i}~,~~
 0.527517 +
  0.299273 \mathbbm{i}~,~~\nonumber\\
  && 0.493798~,~~ -6.80042~,~~ 1.144984~,~~ 0.795994~,~~ -1.69080~,~~
0.0420044~,~~ 0.569967 - 1.110081 \mathbbm{i}~,~~\nonumber\\
&&
 0.569967 +
  1.110081 \mathbbm{i}~,~~ -0.00857861~,~~ 0.212916~,~~ 0.00455376~,~~ 0.0855984~,~~
0.00970033~,~~\nonumber\\
&& 0.0145344\}~,~~~\nonumber\eea}
and
{\footnotesize \bea \diag (T^d_{z_6})& =&\{-64.2332~,~~ 51.9097 +
32.8860 \mathbbm{i}~,~~
 51.9097 - 32.8860 \mathbbm{i}~,~~ -56.7763~,~~ -36.8800 + 1.7486 \mathbbm{i}~,~~\nonumber\\
 && -36.8800 -
  1.7486 \mathbbm{i}~,~~ -12.51570~,~~ 6.22689~,~~ 2.32830 + 1.39061 \mathbbm{i}~,~~
 2.32830 -
  1.39061 \mathbbm{i}~,~~~~~~~~~~~~~~~~~~~~~~~~~~~~~~~~~~~~~~~~~~~~~~~~~~~~~~~~~~~~~~~~~~~~~~~~~\nonumber\\
  && 2.31186~,~~ -1.263937~,~~ 0.712806~,~~ 0.601979~,~~ -0.488528~,~~
0.464830~,~~ 0.154052 + 0.393588 \mathbbm{i}~,~~\nonumber\\
&&
 0.154052 -
  0.393588 \mathbbm{i}~,~~ -0.0892275~,~~ -0.0559545~,~~ 0.0545382~,~~ 0.0445606~,~~
-0.0265855~,~~ \nonumber\\
&&0.00921802\} ~.~~~\nonumber\eea}
It can be checked directly that, each set of diagonal elements
$\{(T_{z_3})_{i,i},(T_{z_4})_{i,i},(T_{z_5})_{i,i},(T_{z_6})_{i,i}\}$
corresponds to a set of solution
$\{z_3^{sol_j},z_4^{sol_j},z_5^{sol_j},z_6^{sol_j}\}$ of scattering
equations. Thus each diagonal element of matrix
$P'(T^d_{z_i})Q'^{-1}(T_{z_i}^d)$ is identical to the integrand
$P/Q$ evaluated at one solution of scattering equations, and the
equivalence between results of these two methods is obvious.


%

With the arithmetic result, it is possible to determine the terms
appearing in amplitude by setting appropriate kinematics. In fact,
in this example, we know that the result should be the sum
\bea
\cA_{7}=\sum_{\I,\I_i}{c_{\I}\over
s_{\I_1}s_{\I_2}s_{\I_3}s_{\I_4}}~,~~~\label{phi3A7candy}\eea
where naively the summation is over all possible products of 4
physical poles $s_{\I_i}$, i.e., $\binom{14}{4}=1001$ terms with
$c_{\I},\I=1,\ldots 1001,$ being either zero or one. By choosing
1001 different group of kinematics for physical poles, we get 1001
linear equations of (\ref{phi3A7candy}), and solving them gives the
$c_{\I}$.

Indeed, the number of terms grows very fast with $n$ in
(\ref{phi3A7candy}). The number of independent poles is
$n'={(n-1)(n-2)\over 2}-1$ for massless theory, while the number of
possible terms in the expansion is $\binom{n'}{n-3}$. For $n=8$, the
number is 15504, and for $n=9$ the number is 296010. So it is not
very doable when $n$ is large. However, for $\phi^3$ theory, the
number of color-ordered diagram is much smaller, and the counting is
${2^{n-2}(2n-5)!!\over (n-1)!}$. So for $n=7$, the possible terms
appearing in (\ref{phi3A7candy}) is 42 (an auspicious number). For
$n=8$, the number is 132, and for $n=9$, the number is 429, etc. If
we restrict to the 42 possible terms in (\ref{phi3A7candy}), then it
is possible to compute the coefficients $c_{\I}$ by choosing one set
of kinematics.

One can let each physical pole be assigned a random prime number, and
compute $\Tr(P'Q'^{-1})$ and then let Mathematica go through all
$2^{42}$ possibilities of $c_{\I}$'s to find the summation
$\sum_{\I=1}^{42}{c_{\I}\over
s_{\I_1}s_{\I_2}s_{\I_3}s_{\I_4}}=\Tr(P'Q'^{-1})$. If the prime
numbers in kinematic variables are distributed randomly in a very large
scale, e.g., primes between 2 to 10000, then usually we can
find one unique solution for $c_{\I}$ in the spirit of Egyptian fractions.
This enables us to do one computation and fix all coefficients.

For example, let us compute
\bea
\cA'_7=\sum_{sol}{z_{12}^2z_{27}^2z_{71}^2\over
|\Phi|^{127}_{127}}{1\over
  z_{12}z_{23}z_{34}z_{45}z_{56}z_{67}z_{71}z_{12}z_{24}z_{45}z_{57}z_{76}z_{63}z_{31}}~.~~~
\eea
With the kinematics shown in \eqref{primen7}, we find the unique decomposition
\bea \Tr(P'Q'^{-1})={284\over 1037296765}={1\over 5\times 61\times
101\times 151}+{1\over 5\times 101\times 151\times 233}~,~~~\eea
which indicates that
\bea A'_7={1\over s_{12}s_{45}s_{67}s_{123}}+{1\over
s_{12}s_{67}s_{123}s_{567}}~,~~~\eea
and agrees with the result given by CHY mapping  rules \cite{Baadsgaard:2015voa, Baadsgaard:2015ifa}.

There is a way to directly determine whether a certain term ${1\over
s_{\I_1}s_{\I_2}s_{\I_3}s_{\I_4}}$ is present in the result or not,
by setting the kinematics $s_{\I_1}=a$, $s_{\I_2}=a^2$,
$s_{\I_3}=a^4$, $s_{\I_4}=a^8$, and others random primes not
equaling to $a$. If this term exists, then the denominator has a
factor $a^{15}$. Again in the $\cA'_7$ example, if we instead set
$s_{12}=5$, $s_{45}=5^2$, $s_{67}=5^4$, $s_{123}=5^8$, then the
result is ${248\over 5^{15}\times 223}$, thus ${1\over
s_{\I_1}s_{\I_2}s_{\I_3}s_{\I_4}}$ is a term in $A'_{7}$. However,
if we set $s_{12}=5$, $s_{56}=5^2$, $s_{67}=5^4$, $s_{123}=5^8$,
then the result is ${284\over 5^{13}\times 61\times 223}$. This
indicates that ${1\over s_{12}s_{56}s_{67}s_{123}}$ is not a term in
$\cA'_7$, while the $5^{13}$ factor indicates that possible terms
involving ${1\over s_{12}s_{67}s_{123}}$ must exist, which provides
further information for detecting other existing terms. By this way,
we can check all possible terms by setting kinematics for each one.

The number of solutions for scattering equations grows as $(n-3)!$,
while the companion matrix grows as $(n-3)!\times (n-3)!$. When
$n=8$, we need to invert the matrix $Q'_{120\times 120}$, and at
$n=9$, the matrix $Q'_{720\times 720}$, etc.
This sets the limitation on the computation of higher $n$.

\section{Conclusions and Outlook}\label{s:conc}

In this paper, motivated by the explanation of equivalence of different
integrands in the CHY setup, we propose a new method using {\bf companion matrices}, borrowed from the study of zero-dimensional ideals in computational algebraic geometry, to evaluate the integrand.
One advantage of the method is that the rationality of final integral is obvious.
Thus although our method may not be as efficient as the one proposed in \cite{Baadsgaard:2015voa, Baadsgaard:2015ifa, Cachazo:2015nwa}, it does give a new angle to study the important problem of scattering amplitudes.

As shown in the plethora of examples, when the number of external legs grows,
the analytic expression of companion matrix becomes harder. In fact, when $n\geq 6$, the best way to do it is by assigning the kinematic variables to random prime numbers in order to reconstruct the analytic result.
The salient feature of our method is that it is purely linear-algebraic, involving nothing more than finding the inverse and trace of matrices.
The linearity of the trace, for example, was demonstrated to immediately lead to non-trivial identities in the amplitudes.

Now, since the physical problem is very symmetric as can be seen by
the polynomials given in \eref{DG-2} and \eref{DG-2-1}, one is confronted with an immediate mathematical challenge.
If we could analytically find, say by induction, the Gr\"obner basis and subsequent monomial basis for the polynomial form of the scattering equations in some appropriate lexicographic ordering, then one would find a recursive way to construct the companion matrix explicitly, much like the recursive construction of tree-level amplitude by using BCFW deformation \cite{Britto:2004ap,Britto:2005fq}.
Working out this construction is hard but worthwhile, as it would give explicit analytic results for the amplitudes and provide a deeper understanding of the CHY formalism.


\section*{Acknowledgments}

We would like to thank Yang Zhang for discussion and sharing the
similar idea of using computational algebraic geometry. We would also like
to thank C.~Baadsgaard, N.~Bjerrum-Bohr, P.~H. Damgaard for the discussion. RH
acknowledges discussions with Qingjun Jin. BF is grateful to the
Qiu-Shi Fund and the Chinese NSF under contracts No.11031005,
No.11135006 and No.11125523; he would like to thank the hospitality
of the Niels Bohr International Academy and City University, London.
YHH would like to thank the Science and Technology Facilities
Council, UK, for grant ST/J00037X/1, the Chinese Ministry of
Education, for a Chang-Jiang Chair Professorship at NanKai
University as well as the City of Tian-Jin for a Qian-Ren
Scholarship, as well as City University, London and Merton College,
Oxford, for their enduring support. RH would also like to thank the
supporting from Chinese Postdoctoral Administrative Committee.



\bibliographystyle{JHEP}
\bibliography{paperEqual2}

\end{document}